\documentclass{jfm}
\include{srctex}
\usepackage{graphicx}
\usepackage{natbib}
\usepackage{color}
\usepackage{amsmath}

\ifCUPmtlplainloaded \else
  \checkfont{eurm10}
  \iffontfound
    \IfFileExists{upmath.sty}
      {\typeout{^^JFound AMS Euler Roman fonts on the system,
                   using the 'upmath' package.^^J}%
       \usepackage{upmath}}
      {\typeout{^^JFound AMS Euler Roman fonts on the system, but you
                   dont seem to have the}%
       \typeout{'upmath' package installed. JFM.cls can take advantage
                 of these fonts,^^Jif you use 'upmath' package.^^J}%
      }
  \else
  \fi
\fi

\ifCUPmtlplainloaded \else
  \checkfont{msam10}
  \iffontfound
    \IfFileExists{amssymb.sty}
      {\typeout{^^JFound AMS Symbol fonts on the system, using the
                'amssymb' package.^^J}%
       \usepackage{amssymb}%
       \let\le=\leqslant  \let\leq=\leqslant
       \let\ge=\geqslant  \let\geq=\geqslant
      }{}
  \fi
\fi

\ifCUPmtlplainloaded \else
  \IfFileExists{amsbsy.sty}
    {\typeout{^^JFound the 'amsbsy' package on the system, using it.^^J}%
     \usepackage{amsbsy}}
    {\providecommand\boldsymbol[1]{\mbox{\boldmath $##1$}}}
\fi

\providecommand\bnabla{\boldsymbol{\nabla}} 
\providecommand\bcdot{\boldsymbol{\cdot}} 

\newcommand\Rey{\mbox{\textit{Re}}}  

\newsavebox{\astrutbox}
\sbox{\astrutbox}{\rule[-5pt]{0pt}{20pt}}

\newcommand\beq{\begin{equation}}
\newcommand\eeq{\end{equation}}
\newcommand\beqa{\begin{eqnarray}}
\newcommand\eeqa{\end{eqnarray}}
\newcommand{\nn}{\nonumber\\}
\newcommand{\dd}{\text{d}}
\newcommand{\NS}{\text{NS}}

\title[Steady base states for granular hydrodynamics]{Steady  base states for
 non-Newtonian granular hydrodynamics}

\author[{F. Vega Reyes}, {A. Santos} and {V. Garz\'o}]%
{F\ls R\ls A\ls N\ls C\ls I\ls S\ls C\ls O\ns V\ls E\ls G\ls A\ns R\ls E\ls Y\ls E\ls S%
\thanks{Email address for correspondence: fvega@unex.es}, \ns A\ls N\ls D\ls R\ls \'E\ls S\ns S\ls A\ls N\ls T\ls O\ls S
\and \ns V\ls I\ls C\ls E\ls N\ls T\ls E\ns G\ls A\ls R\ls Z\ls \'O\break
}

\affiliation{Departamento de F\'isica, Universidad de Extremadura, 06071 Badajoz, Spain}

\date{\today}
\begin{document}

\maketitle

\begin{abstract}
We study in this work steady laminar flows in a low density granular gas
modelled as a system of identical smooth hard spheres that collide
inelastically. The system is excited by shear and temperature sources at the
boundaries, which consist of two infinite parallel walls. Thus, the geometry of
the system is the same that yields the planar Fourier and Couette flows in
standard gases. We show that it is possible to describe the steady granular
flows in this system, even at large inelasticities, by means of a
(non-Newtonian) hydrodynamic approach. All five types of Couette--Fourier granular
flows are systematically described, identifying the different types of
hydrodynamic profiles. Excellent agreement is found between our classification of flows
and simulation results. Also, we obtain the corresponding non-linear transport
coefficients by following three independent and
complementary methods: (1) an analytical solution obtained from Grad's
13-moment method applied to the inelastic Boltzmann equation, (2) a numerical
solution of the inelastic Boltzmann equation obtained by means of the direct
simulation Monte Carlo method and (3) event-driven molecular dynamics
simulations. We find that, while Grad's theory does not describe quantitatively well all transport coefficients, the three procedures yield the same general classification of planar Couette--Fourier flows for the granular gas.
\end{abstract}


\section{Introduction}
\label{intro}

There have been in the recent years a large number of studies on the dynamics of granular gases, where `granular gas' is a term used to refer to a low density system of many mesoscopic particles that collide inelastically in pairs. {Due} to inelasticity in the collisions, the granular gas particles tend to collapse to a rest state, unless there is some kind of energy input. In particular, \citet[]{GZ93} showed that clustering instabilities spontaneously appear in a freely evolving granular gas. Nevertheless,  most situations of practical interest involve an energy input to compensate for the energy loss and sustain, in some cases, the `gas' condition of the granular system. This type of problem has been extensively studied, giving rise to a subfield of granular dynamics: `rapid granular flows' \citep*{JS83,WJS96,G03,AT06}. Furthermore, it has been shown that rapid granular flows can attain steady states, some of which, under appropriate circumstances and for simple geometries, can give rise to laminar flows, in the same way as a regular gas does \citep*[see, for instance, the work by][on Couette granular flows]{TTMGSD01}. The question arising \citep{G03} is, what is the appropriate theoretical approach to study these granular flows?

Let us start with classical non-equilibrium statistical mechanics for an ideal gas described by the Boltzmann equation \citep{CC70}. As is well known,  the equilibrium velocity distribution function $f(\boldsymbol{r},\boldsymbol{v},t)$ for an {ordinary (i.e., elastic)} gas is the Maxwell--Boltzmann distribution  \citep{H87}.  For non-equilibrium states, however, the solution  of the Boltzmann equation is generally not known. On the other hand, in some cases, there exist special solutions where all the space and time dependence of $f(\boldsymbol{r},\boldsymbol{v},t)$ occurs only through a functional dependence on the average fields {$n$ (density), $\boldsymbol{u}$ (flow velocity) and $T$ (temperature)} associated with the conserved quantities (mass, momentum and energy)
\citep[]{CC70}. This type of solution is called a \emph{normal} solution of the Boltzmann equation \citep{C88}. As a consequence, the momentum and heat fluxes are also functionals of the hydrodynamic fields  and thus the balance equations become a closed set of  equations for those fields.
Therefore, the \textit{normal} solutions of the Boltzmann equation yield a \textit{hydrodynamic} description \citep{H83}, since the closed set of equations is
actually formally similar to the traditional fluid mechanics equations \citep{CC70}. In practice, what we have got
is a transition from a microscopic description (based on the distribution function) to a
macroscopic description (based on the average fields) \citep{H12}.

When the strength of the hydrodynamic gradients is small, the above functional dependence of the non-uniform distribution function $f$ on $n, \boldsymbol{u}, T$ can be constructed by means of the Chapman--Enskog method \citep{CC70}, whereby
$f$ is expressed as a series in a formal parameter $\epsilon$:
\beq
f=f^{(0)}+f^{(1)}\epsilon+f^{(2)}\epsilon^2+f^{(3)}\epsilon^3+\cdots.
\eeq
The parameter $\epsilon$ indicates the order in the spatial gradients of the average
fields, scaled with the inverse of a typical microscopic length unit (mean free path, for instance).
If terms up to only first order in the gradients are considered
($f\simeq f^{(0)}+f^{(1)}\epsilon$), the mass, momentum and energy balance equations  are
the well known Navier--Stokes (NS) equations of fluid mechanics \citep{CC70,C88}.
This approach is accurate for problems where the spatial gradients are sufficiently small.
For not so small gradients, terms up to second order in the gradients need to be considered, and we obtain the Burnett equations \citep{B35}, used for instance {in} rarefied gases \citep*{MHGS98,MHSG99,AYB01}.
For both NS and Burnett equations, the expressions
for the fluxes
include a set of parameters called `hydrodynamic transport coefficients'.

Regarding the granular gas, and from a theoretical point of view, it makes sense in principle, due to the system's low density, to derive the dynamics from a closed kinetic equation for the distribution function of a single particle, in an analogous way to the standard gas \citep{G03}; i.e., it is  assumed that pre-collisional velocities are not statistically correlated (or, at least, that their correlations are not important). Thus, the corresponding kinetic equation is analogous to the Boltzmann equation but with the modification that inelasticity introduces in the collision integral part {\citep*{BDKS98,G03}}. We may call this modified version of the Boltzmann equation  `inelastic Boltzmann equation' \citep{BDKS98,G03}. In addition, if we assume the existence of a normal solution to the inelastic Boltzmann equation, a hydrodynamic description  analogous to that described above for an elastic gas results for a granular gas; i.e., transport coefficients and a set of hydrodynamic equations may be derived.
This is obviously a question of much interest in the description of transport properties of large sets of grains at low density.

However, due to the coupling between spatial gradients and inelasticity in steady states {\citep*{SG98,SGD04}}, the collisional cooling sets the strength of the spatial gradients and thus scale separation might not occur (i.e., gradients might not be small), except in the limit of quasi-elastic collisions \citep{VU09}. Therefore, NS or Burnett hydrodynamics would only be expected to work well for steady granular flows in the quasi-elastic limit. Nevertheless, some recent works have found that a \emph{non-Newtonian} hydrodynamic description of planar laminar flows, beyond Burnett order, is still possible for moderately large spatial gradients, even for large inelasticity \citep*{TTMGSD01,SGV09,VSG10,VGS11}. Actually, it is not surprising that a generalized hydrodynamic description of the Boltzmann inelastic equation works in rapid granular flows, even for moderately large gradients, since this is also possible when strong gradients occur in elastic gases \citep{AYB01,GS03}. We have pointed out previously that this implies that hydrodynamics for granular gases is a generalization of classic hydrodynamics for elastic gases. Furthermore, a special class of flows has been recently found in a unified hydrodynamic description valid for elastic and inelastic gases {\citep{VSG10,VGS11}}. Thus, the only formal difference between transport theory for granular and ordinary gases would emerge not from the limitations due to scale separation but from the possible influence of statistical correlations arising from memory effects due to inelasticity. In fact, there is a number of works showing velocity correlations in systems of inelastic particles \citep*[for instance, see the work by][]{ML98,SM01,SPM01,PTNE02,PEU02,BPKZ07} and elastic particles \citep{SH07}.
This statistical effect would have its origin at the more fundamental level of the kinetic equation (the inelastic Boltzmann equation). Put in other words, if the Boltzmann inelastic equation is to be valid, hydrodynamic solutions for steady granular flows arising from it should work, as  has been previously shown  by different authors \citep{AN98,TTMGSD01,VSG10}.  As a matter of fact, the inelastic Boltzmann  equation has been used, with good results, as the starting point in an overwhelming number of studies on rapid granular flows \citep{G03,AT06}. Additionally, good agreement has also been shown,  for a variety of rapid granular flows, between hydrodynamic theory (stemming from the inelastic Boltzmann equation) and molecular dynamics results  {\citep*[in which the velocity statistical correlations would be inherently present, see the works by][]{PEU02,LBD02,DHGD02,AL03,MGAL06}}. Furthermore, in the case of the special class mentioned before, the agreement of molecular dynamics results with (Grad's) hydrodynamic theory is excellent \citep{VSG10,VGS11}.

A considerable amount of work has been devoted to systematic {calculations} of hydrodynamic transport coefficients for granular gas systems, with different degrees of approach in the perturbative solution of the non-uniform distribution function {\citep{SG98,BDKS98,G03,NAAJS99,AANGH05}}. However, the derivation of {non-Newtonian transport coefficients in} simple laminar flows has been probably not as systematic as for the case of NS transport coefficients.

\begin{figure}
\begin{center}
\includegraphics[width=0.8\columnwidth]{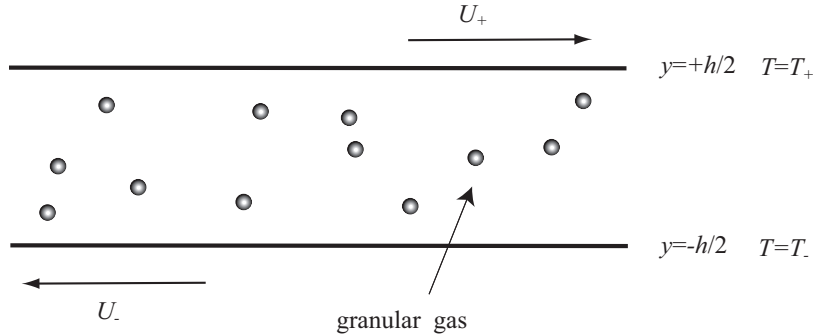}
\end{center}
\caption{Schematic view of the system subject of study. The granular gas is heated and sheared from two infinite parallel walls. Walls are located at $y=\pm h/2$ and have temperatures $T_\pm$ and velocities $U_\pm$, respectively.} \label{fig1}
\end{figure}

The main goal of this paper is the systematic derivation, by means of a non-Newtonian hydrodynamic approach, of the steady profiles for laminar granular flows in the {simple} geometry of two infinite parallel walls containing the gas. More specifically, shear and energy are input from the walls  (see figure \ref{fig1}). In the theoretical approach we assume that (i) the hydrostatic pressure $p$ is constant, (ii) the \emph{reduced} shear rate $a$ (i.e., the  ratio between the local shear rate and the local collision frequency) is also constant, (iii) the shear stress is independent of the granular temperature gradient $\partial_y T$, whereas (iv) the heat flux $q_y$ is proportional to $\partial_y T$.
As we will see, the resulting classification of profiles is formally analogous  to the {one} found for NS hydrodynamics in the quasielastic limit \citep{VU09}, except that the constitutive relations are non-linear. This classification  is done based on the signs of $\partial_y(T^{1/2}\partial_y T)$ and $\partial_y^2 T$.  As we will show, both signs remain constant throughout the system {and} are related to the competition between viscous heating and inelastic cooling. Moreover, the sign of $\partial_y^2 T$ is also governed by the wall temperature difference. In the case of elastic collisions, only the {viscous heating} effect is present and so $\partial_y(T^{1/2}\partial_y T)<0$, which implies $\partial_y^2T<0$ \citep{GS03}. Therefore, the general classification is only relevant for granular gases and, consequently, the case of ordinary gases is embedded as a particular case.

\begin{figure}
\begin{center}
\includegraphics[width=0.7\columnwidth]{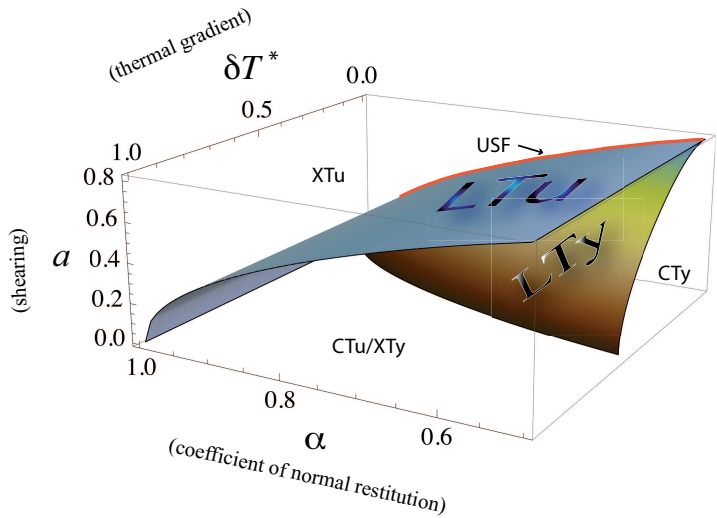}
\end{center}
\caption{Each point of this diagram represents a steady-state Couette--Fourier flow defined univocally by the set of parameters $\delta T^*$ (difference between the temperatures at the top and bottom fluid layers, divided by the wall separation), $a$ (reduced shear rate) and $\alpha$ (coefficient of restitution), the two first ones being determined from the boundary conditions. The surface with the label LTu defines the class of states where the temperature $T$ is a linear function of the flow velocity $u_x$, while the surface labelled as LTy (below the LTu surface)  defines the class with a linear profile $T(y)$. Both surfaces intersect in the line representing the uniform shear flow (USF), located in the $\delta T^*=0$ plane. In addition, the LTu surface contains the line corresponding to Fourier flows for ordinary gases (represented by the $\delta T^*$ axis, i.e., $a=0$ and $\alpha=1$). The point $\delta T^*=0$, $a=0$ and $\alpha=1$ (not visible in the diagram) represents the equilibrium state of an ordinary gas.  Notice that, whereas the LTu surface has points for all values of $\delta T^*$, the LTy surface has an upper bound of $\delta T^*$  which occurs at $a=0$ for each $\alpha$. The LTu and LTy surfaces split the space into  three regions: XTu, CTu/XTy and CTy (see
\S~\ref{class.3}).} \label{LTyLTu}
\end{figure}

The hypotheses (i)--(iv) are  {sensible for a number of reasons. First,} they have shown a good agreement with computer simulations in previous works on Couette granular gas flows in the particular case $\partial_y(T^{1/2}\partial_y T)<0$ \citep{TTMGSD01}.
In addition, there exists a special class of flows, including both elastic and inelastic flows {\citep{VU09,SGV09,VSG10,VGS11}}, characterized by {$\partial_y(T^{1/2}\partial_y T)=0$. This special class} defines a surface in the {three}-parameter space conformed by inelasticity (represented by the coefficient of normal restitution $\alpha$), reduced shear rate and thermal gradient, {as shown in figure \ref{LTyLTu}. It} is called `LTu' surface since this class of flows is characterized by having linear $T(u_x)$ profiles \citep{VSG10,VGS11}. The LTu surface splits the  parameter space into two regions: the first region (above the LTu surface in figure \ref{LTyLTu} {and labelled XTu}) corresponds to $\partial_y(T^{1/2}\partial_y T)<0$ (i.e., viscous heating  overcomes inelastic cooling), while the second region (below the LTu surface) has $\partial_y(T^{1/2}\partial_y T)>0$ (i.e., inelastic cooling dominates).
As we will see, the region below the LTu surface can also be split into two {sub-}regions {(labelled CTu/XTy and CTy)}, depending on the sign of $\partial_y^2 T$, separated by a surface where $\partial_y^2 T=0$. The latter surface is called here {`LTy'}  because it corresponds to states where $T(y)$ is a linear function.
To the best of our knowledge, the {regions} below the LTu surface {have} not been explored before for $a\neq0$, except in the NS description \citep{VU09}. All other studies below the LTu surface have been restricted to the plane $a=0$ in figure \ref{LTyLTu} {\citep*[see, for instance, the works by][]{GZB97,BC98,BRM00}. The most prominent result in studies for the $a=0$ plane is perhaps the finding of LTy states \citep*{BCMR01,BKR09,BKD11,BKD12}}, which are represented in figure \ref{LTyLTu} by  the intersection curve between the LTy surface and the plane $a=0$.

Our purpose is now to extend results obtained in  previous works by providing a comprehensive description of  granular/elastic Couette--Fourier gas flows, as depicted in figure \ref{LTyLTu}. {For instance, by determination of the LTy surface we get to connect the LTy states for $a=0$ found by \citet{BCMR01} with the well known uniform shear flow \citep[USF, also referred to as `simple shear flow', see for instance the works by][]{C89}, within the same theoretical frame}. We will follow three complementary routes. First, we will undertake a theoretical description based on Grad's 13-moment method \citep{G49}. Second, we will obtain results from two independent simulation methods, the direct simulation Monte Carlo (DSMC) method, from which a numerical solution of the inelastic Boltzmann equation is obtained, and event-driven molecular dynamics (MD) simulations, which solve Newton's equations of inelastic hard spheres.
As we will show, both simulation techniques support the classification of states mentioned before (and sketched in figure \ref{LTyLTu}). Moreover, the non-Newtonian transport coefficients obtained from the approximate Grad solution agree reasonably well with simulations.

The structure of this work is as follows. In \S~\ref{system} we describe in more detail the system under study and write the corresponding kinetic and average balance equations. For the sake of completeness, the solution at the NS level is briefly recalled in \S~\ref{secNS}. Next,
the theoretical Grad's solution is derived in  \S~\ref{grad}. In \S~\ref{class} the assumptions (i)--(iv) referred to above are introduced and the associated classification of states is worked out.  In \S~\ref{results} we briefly describe the computational methods and compare the simulation results with Grad's theory. Finally, we conclude the paper with a {summary and } discussion in \S~\ref{conclusions}.

\section{Boltzmann kinetic theory and general balance equations}
\label{system}

The system we study is depicted in figure \ref{fig1}. It is bounded by two infinite parallel walls from where we input energy to a granular gas enclosed in between. The energy is input by heating (both walls are in general at different temperatures) and, optionally, shearing (walls may be moving at different velocities). The granular gas is composed by a large number of inelastic smooth hard disks/spheres (inelastic because kinetic energy is not conserved during collisions). We consider a set of disks/spheres that is sufficiently sparse at all times; i.e., the rate at which energy is input is always intense enough so that kinetic energy loss in collisions will not cause the system to `freeze' or `collapse' {\citep*[so `inelastic collapse' does not occur; see for instance][]{GZ93,KLM10}}. By sufficiently sparse we mean that we deal with a \textit{gas} in the kinetic theory sense: collisions are only binary and instantaneous (time during collisions is very short compared to typical time between consecutive collisions). We consider also that their pre-collision velocities are statistically uncorrelated (`molecular chaos' assumption). Therefore, in the absence of external forces, we will assume that the velocity distribution function of the system obeys the inelastic Boltzmann kinetic equation \citep{BDKS98,BP04}
\beq
\left(\frac{\partial}{\partial t}
+\boldsymbol{v}\bcdot\bnabla\right)f(\boldsymbol{r},\boldsymbol{v};t)=J[\boldsymbol{v}|f,f],
\label{BE1}
\eeq
with $J$ being the collisional integral, whose expression is
\beqa
J\left[\boldsymbol{v}_1|f,f\right] &=&\sigma^{d-1} \int
\dd\boldsymbol{v}_{2}\int \dd\widehat{\boldsymbol{\sigma}} \, \Theta\left(\boldsymbol{g}\bcdot \widehat{\boldsymbol{\sigma}}\right)\left(\boldsymbol{g}\bcdot \widehat{\boldsymbol{\sigma}}\right)\left[\alpha^{-2}f(\boldsymbol{r},\boldsymbol{v}'_1;t)f(\boldsymbol{r},\boldsymbol{v}_{2}';t)\right.
\nn
&&\left.
-f(\boldsymbol{r},\boldsymbol{v}_1;t)f(\boldsymbol{r},\boldsymbol{v}_{2};t)\right] ,
\label{BE}
\eeqa
where $d$ is the dimensionality, $\sigma$ is the diameter of a sphere, $\Theta(x)$ is Heaviside's step function, $\widehat{\boldsymbol{\sigma}}$ is a unit vector directed along the line joining the centers of the colliding pair, $\boldsymbol{g}=\boldsymbol{v}_1-\boldsymbol{v}_2$ is the relative velocity, and $\{\boldsymbol{v}_1, \boldsymbol{v}_{2}\}$ and $\{\boldsymbol{v}_1^{\prime}, \boldsymbol{v}_{2}^{\prime}\}$ are post-collisional and pre-collisional velocities respectively. As we see in (\ref{BE}), $J[\boldsymbol{v}_1|f,f]$ depends on the parameter $\alpha$, {which} characterizes inelasticity in the collisions and is called coefficient of normal restitution \citep{BDKS98,G03}. The (restituting) collisional rules for a pair of colliding inelastic smooth hard disks/spheres is
\beqa
\boldsymbol{v}_1^{\prime}&=&\boldsymbol{v}_1-\frac{1}{2}\left( 1+\alpha
^{-1}\right)(\widehat{\boldsymbol{\sigma}}\bcdot \boldsymbol{g})\widehat{\boldsymbol {\sigma}}, \nn
\boldsymbol{v}_{2}^{\prime}&=&\boldsymbol{v}_{2}+\frac{1}{2}\left( 1+\alpha^{-1}\right)
(\widehat{\boldsymbol{\sigma}}\bcdot \boldsymbol{g})\widehat{\boldsymbol{\sigma}}.
\eeqa

The first $d+2$ velocity moments of $f(\boldsymbol{r},\boldsymbol{v},t)$ define the number density $n(\boldsymbol{r},t)$, the flow velocity $\boldsymbol{u}(\boldsymbol{r},t)$ and the granular temperature $T(\boldsymbol{r},t)$ as
\beq
n=\int \dd\boldsymbol{v}\,f(\boldsymbol{v}),
\label{n}
\eeq
\beq
n\boldsymbol{u}=\int \dd\boldsymbol{v}\,\boldsymbol{v}f(\boldsymbol{v}),
\label{u}
\eeq
\beq
nT={\frac{m}{d}}\int \dd\boldsymbol{v}\,V^2f(\boldsymbol{v}),
\label{T}
\eeq
where $\boldsymbol{V}\equiv \boldsymbol{v}-\boldsymbol{u}$ is the peculiar velocity and $m$ is the mass of a particle.

Mass, momentum and energy balance equations are obtained by multiplying both sides of \eqref{BE1} by $1$, $\boldsymbol{v}$, $v^2$ and integrating over velocity. The results are
\begin{equation}
{D_tn}=-n\bnabla\bcdot\boldsymbol{u} ,
\label{nbal}
\end{equation}
\begin{equation}
{D_t\boldsymbol{u}}=-\frac{1}{mn}\bnabla\bcdot{\mathsfbi{P}}  ,
\label{Pbal}
\end{equation}
\begin{equation}
{D_tT}+\zeta T=-\frac{2}{dn}\left(\mathsfbi{P}\boldsymbol{:\bnabla}\boldsymbol{u}+\bnabla\bcdot{\boldsymbol q}\right)  .
\label{Tbal}
\end{equation}
In the above equations, $D_t\equiv \partial_t+\boldsymbol{u}\bcdot\bnabla$ is the material derivative,
\beq
\mathsfbi{P}=m\int \dd\boldsymbol{v}\,\boldsymbol{V}\boldsymbol{V}f(\boldsymbol{v})
\label{P}
\eeq
is the pressure tensor,
\beq
\boldsymbol{q}=\frac{m}{2}\int \dd\boldsymbol{v}\,V^2\boldsymbol{V}f(\boldsymbol{v})
\label{q}
\eeq
is the heat flux vector and
\beq
\zeta=-\frac{m}{dnT}\int \dd\boldsymbol{v}\,v^2 J[\boldsymbol{v}|f,f]
\label{zeta}
\eeq
is the cooling rate characterizing the rate of energy dissipated due to collisions.

Next, we consider the steady base states that may be generated from energy input in our geometry. Independently of the nature of the boundary conditions, and if there is no pressure drop source or gravitational field in the horizontal directions
\citep[which may generate Poiseuille flows; see for example the recent works by][]{TS04,ST06,AC10}, the spatial dependence of these steady base states will occur only in the coordinate $y$, perpendicular to both walls (we call it vertical direction). Moreover, the flow velocity is expected to be parallel to the walls, i.e., $\boldsymbol{u}(y)=u_x(y)\boldsymbol{e}_x$. Consequently, the Boltzmann equation \eqref{BE1} for these reference steady states can be rewritten as
\beq
v_y\frac{\partial f}{\partial y}=J[f,f]
\label{EB2}
\eeq
and the balance equations have the simple forms
\beq
\frac{\partial P_{xy}}{\partial y}=0, \quad \frac{\partial P_{yy}}{\partial y}=0, \label{Pct}
\eeq
\beq
-\frac{2}{dn}\left(P_{xy}\frac{\partial u_x}{\partial y}+\frac{\partial q_y}{\partial y}\right)=\zeta T.
\label{balance}
\eeq
Due to the symmetry of the problem, all the off-diagonal elements of the pressure tensor different from $P_{xy}$ vanish {and, in principle, the two shear-flow plane diagonal elements ($P_{xx}$ and $P_{yy}$) are different whereas the remaining $d-2$ diagonal elements orthogonal to the shear-flow plane are equal}. The latter property implies that $P_{xx}+P_{yy}+(d-2)P_{zz}=dp$, where  $p=nT=d^{-1}\text{Tr}\mathsfbi{P}$ is the hydrostatic pressure.

\section{Navier--Stokes description}
\label{secNS}
The balance equations \eqref{Pct} and \eqref{balance} are exact and do not assume any particular form for the constitutive equations. However, they do not constitute a closed set of equations for the hydrodynamic fields.

The simplest  approach to close the problem is provided by the NS constitutive equations, which, in the geometry of the planar Couette--Fourier flow read \citep{BDKS98,BC01}
\beq
P_{xx}=P_{yy}=P_{zz}=p,
\label{NS1}
\eeq
\beq
P_{xy}=-\eta_0\eta_\NS^*(\alpha)\frac{\partial u_x}{\partial y},
\label{NS2}
\eeq
\beq
q_x=0,
\label{NS3}
\eeq
\beq
q_y=-\lambda_0\kappa_\NS^*(\alpha)\frac{\partial T}{\partial y}-\lambda_0\frac{T}{n}\mu_\NS^*(\alpha)\frac{\partial n}{\partial y}.
\label{NS4}
\eeq
In equations \eqref{NS2} and \eqref{NS4},
\beq
\eta_0=\sqrt{{mT}}{c_\eta}\Lambda_d\sigma^{-(d-1)},\quad
\Lambda_d\equiv\frac{d+2}{8}\Gamma (d/2)\pi^{-\frac{d-1}{2}},
\label{eta0}
\eeq
is the NS shear viscosity for elastic gases \citep{G49,CC70} and
\beq
\lambda_0=\frac{d(d+2)}{2(d-1)}{\frac{c_\lambda}{c_\eta}}\frac{\eta_0}{m}
\label{lambda0}
\eeq
is the NS thermal conductivity for elastic gases \citep{G49,CC70}.
{In equations \eqref{eta0} and \eqref{lambda0}, the factors $c_\eta$ and $c_\lambda$ take the values $c_\eta=1.022$, $c_\lambda=1.029$ for hard disks ($d=2$) and $c_\eta=1.016$, $c_\lambda=1.025$ for hard spheres ($d=3$) \citep{B35,CC70}.}
Finally, $\eta_\NS^*$, $\kappa_\NS^*$ and $\mu_\NS^*$ are the reduced NS transport coefficients of a dilute granular gas, whose expressions are given in Appendix \ref{appNS}. In equations \eqref{etaNS}--\eqref{muNS},
\beq
\zeta^*(\alpha)=\frac{d+2}{4d}(1-\alpha^2)
\label{3.7}
\eeq
represents the ratio between the cooling rate $\zeta$ and an effective collision frequency defined as
\beq
\nu\equiv \frac{p}{\eta_0}.
\label{nu0}
\eeq
Note that $\nu\propto n T^{1/2}$ and thus it depends on $y$.

Now we combine the NS constitutive equations with the three balance equations \eqref{Pct} and \eqref{balance}. First, the exact property $P_{yy}=\text{const}$, together with equation \eqref{NS1}, implies that the hydrostatic pressure is uniform. Next, the exact property $P_{xy}=\text{const}$, together with equation \eqref{NS2}, implies that the product $\eta_0\partial u_x/\partial y=\text{const}$. These two implications can be combined into $a=\text{const}$, where
\begin{equation}
a\equiv\frac{1}{\nu}\frac{\partial u_x}{\partial y}
\label{udy}
\end{equation}
is the \emph{reduced} shear rate.
Finally, we consider the energy balance equation \eqref{balance}. First, since $p=\text{const}$, equation \eqref{NS4} can be rewritten as
\beq
q_y=-\lambda_0\lambda_\NS^*(\alpha)\frac{\partial T}{\partial y},\quad \lambda_\NS^*=\kappa_\NS^*-\mu_\NS^*.
\label{NS5}
\eeq
Next, using the properties $P_{xy}=\text{const}$, $p=\text{const}$ and $a=\text{const}$ in equation \eqref{balance}, one has $\nu^{-1}\partial q_y/\partial y=\text{const}$. This, together with equation \eqref{NS5} yields

\begin{equation}
\frac{1}{\nu}\frac{\partial }{\partial y }\left(\frac{1}{\nu}\frac{\partial T}{\partial y}\right)=-2m\gamma_\NS(\alpha,a),
\label{nuTdy0}
\end{equation}
where
\begin{equation}
\label{gammaNS}
\gamma_\NS(\alpha,a)\equiv\frac{d-1}{d(d+2)}\frac{\eta_\NS^*(\alpha)a^2-\frac{d}{2}\zeta^*(\alpha)}{\lambda_\NS^*(\alpha)}.
\end{equation}

Therefore, the NS description, as applied to the Couette--Fourier flow, predicts that the hydrostatic pressure $p=nT$, the reduced shear rate \eqref{udy} and the second order derivative $(\nu^{-1}\partial_y)^2T$ are uniform. A detailed account of  this NS description was presented by \cite{VU09}.

\section{Non-Newtonian description: Grad's 13 moment method}
\label{grad}
The results derived in \S~\ref{secNS} are restricted to small spatial gradients. Thus, they do not capture non-Newtonian effects, such as normal stress differences (i.e., $P_{xx}\neq P_{yy}\neq P_{zz}$) and a non-zero component of the heat flux orthogonal to the thermal gradient (i.e., $q_x\neq 0$).
Those effects are expected to be present in the solution of the Boltzmann equation beyond the quasi-elastic limit \citep{SG98}.

The aim of this section is to unveil those non-Newtonian properties by solving the set of moment equations derived from the Boltzmann equation by Grad's 13-moment method {\citep{G49}}. In this method, the velocity distribution function $f$ is approximated by the form
\beq
f\to f_0 \left\{1+\frac{m}{2nT^2}\left[\left(P_{ij}-p\delta_{ij}\right)V_iV_j+\frac{4}{d+2}\left(
\frac{mV^2}{2T}-\frac{d+2}{2}\right)\boldsymbol{V}\cdot \boldsymbol{q}\right]\right\},
\label{3.1}
\eeq
where
\beq
f_0=n\left(\frac{m}{2\pi T}\right)^{d/2} e^{-mV^2/2T}
\label{3.2}
\eeq
is the local equilibrium distribution. The number of moments involved in equation \eqref{3.1} is $d(d+5)/2+1$, which becomes 13 in the three-dimensional case. The coefficients in Grad's distribution function have been
obtained by requiring the pressure tensor and heat flux of the trial function (\ref{3.1})
to be the same as those of the exact distribution $f$.

{The Grad distribution \eqref{3.1} can be interpreted as the linearization of the maximum-entropy distribution constrained by the first $d(d+5)/2+1$ moments \citep{K10a}. {}From that point of view, it is not guaranteed \emph{a priori} that it is quantitatively accurate for large deviations from the local equilibrium distribution. Moreover, an extra isotropic term associated with the fourth velocity moment can also be included \citep{SG98}. However, here we consider the \emph{minimal}  version of Grad's method,  restricting the number of non-Maxwellian parameters to the stress tensor and the heat flux vector, since  extra terms do not significantly increase accuracy.
}

According to the approximation \eqref{3.1}, one has
\beq
\frac{m}{2}\int\dd\boldsymbol{v}\,V_iV_jV_k f\to\frac{1}{d+2}\left(q_i\delta_{jk}+q_j\delta_{ik}+q_k\delta_{ij}\right),
\label{GG1}
\eeq
\beq
\frac{m}{2}\int\dd\boldsymbol{v}\,V^2 V_i V_j f\to \frac{p}{nm}\left(\frac{d+4}{2} P_{ij}-p\delta_{ij}\right).
\label{GG2}
\eeq
In addition \citep{BDKS98,BC01,GM02,VGS11},
\begin{equation}
\label{3.3}
m\int \dd\boldsymbol{v}\,V_iV_jJ[f,f]\to-\nu\left[\beta_1 \left(P_{ij}-p\delta_{ij}\right)+\zeta^* P_{ij}\right],
\end{equation}
\begin{equation}
\label{3.4x}
\frac{m}{2}\int \dd\boldsymbol{v}\,V^2\boldsymbol{V}J[f,f]\to-\nu\frac{d-1}{d}\beta_2 \boldsymbol{q},
\end{equation}
where, as usual,  terms {non-linear} in $P_{ij}-p\delta_{ij}$ and $\boldsymbol{q}$ have been neglected. On the other hand, the quadratic  terms have been  retained in some {other} works \citep{HH82,TK95}. In equations \eqref{3.3} and \eqref{3.4x}, {the collision frequency $\nu$ is given by {\eqref{nu0} (and taking into account equation \eqref{eta0}) with $c_\eta=1$. Also}, $\zeta^*\equiv\zeta/\nu$,} $\beta_1$ and $\beta_2$ are given by equations \eqref{3.7}, \eqref{3.6} and \eqref{3.8}, respectively.

The relevant moments in our system are $p$, $T$, $u_x$, $P_{xy}$, $P_{xx}$, $P_{yy}$, $q_x$ and $q_y$. The exact balance equations  \eqref{Pct} and \eqref{balance} are recovered by multiplying both sides of equation \eqref{EB2} by $V_x$, $V_y$ and $V^2$ and integrating over velocity. In order to have a closed set of differential equations, we need  five additional equations, which are obtained by multiplying both sides of equation  \eqref{EB2} by $V_x V_y$, $V_x^2$, $V_y^2$, $V^2V_x$ and $V^2V_y$ and applying the approximations \eqref{GG1}--\eqref{3.4x}. The results are
\beq
\frac{2}{d+2}\partial_s  q_x+P_{yy}{\partial_s u_x}
=-\left(\beta_1+\zeta^*\right)P_{xy},
\label{3.9}
\eeq
\beq
\frac{2}{d+2}\partial_s  q_y+2P_{xy}{\partial_s u_x}
=-\beta_1\left(P_{xx}-p\right)-\zeta^* P_{xx},
\label{3.10}
\eeq
\beq
\frac{6}{d+2}\partial_s  q_y
=-\beta_1\left(P_{yy}-p\right)-\zeta^* P_{yy},
\label{3.11}
\eeq
\begin{equation}
\label{3.12}
\frac{d+4}{2}\partial_s\left(\frac{T}{m}P_{xy}\right)+\frac{d+4}{d+2} q_y\partial_s u_x=-\frac{d-1}{d}\beta_2
q_x,
\end{equation}
\begin{equation}
\label{3.13}
\partial_s\left[\frac{T}{m}\left(\frac{d+4}{2}P_{yy}-p\right)\right]+\frac{2}{d+2}
q_x\partial_s u_x=-\frac{d-1}{d}\beta_2 q_y,
\end{equation}
where {we have introduced the spatial scaled variable $s(y)$ by}
\beq
{\dd s=\nu(y) \dd y.}
\label{ds}
\eeq
{Note that $\dd s/\sqrt{2T(y)/m}$ measures the elementary vertical distance $\dd y$ in units of the (nominal) mean free path $\sqrt{2T(y)/m}/\nu(y)$ . Therefore, the scaled variable $s(y)$ has dimensions of  speed. Its  limit values are deduced from integration of (\ref{ds}), taking into account that the limit values of $y$ are $y=\pm h/2$}.

It must be stressed that in equations {\eqref{3.9}--\eqref{3.13}} the only assumptions made are the stationarity of the system, the geometry and symmetry properties of the planar Couette--Fourier  flow and the applicability of  Grad's method.

{The exact momentum balance equations} \eqref{Pct} {imply} that $P_{xy}=\text{const}$ and $P_{yy}=\text{const}$. {Moreover, if} one  \emph{assumes} that  $p=\text{const}$, equation \eqref{3.11} yields $\partial_s q_y=\text{const}$. Next, the {exact} energy balance equation \eqref{balance} implies that the reduced shear rate $a=\partial_s u_x$ defined by equation \eqref{udy} is also constant {(recall that  $\zeta^*\equiv\zeta/\nu=\text{const}$)}. Taking all of this into account,   we get that $\partial_s q_x=\text{const}$ and $P_{xx}=\text{const}$ from equations \eqref{3.9} and \eqref{3.10}, respectively.  {Finally, equations  \eqref{3.12} and \eqref{3.13} imply that both $q_x$ and $q_y$ are \emph{proportional} to the thermal gradient $\partial_s T$. As a consequence, $\partial_s^2 T=\text{const}$.}

Since  the {pressure $p$, the shear stress $P_{xy}$ and the shear rate $a=\nu^{-1}\partial_y u_x$  are constant, it follows that the  ratio $P_{xy}/\eta_0\partial_y u_x$ is also constant (recall that $\eta_0=p/\nu$). That ratio defines  a (reduced) non-Newtonian shear viscosity coefficient $\eta^*(\alpha, a)$ by}
\beq
P_{xy}=-\eta_0\eta^*(\alpha, a)\frac{\partial u_x}{\partial y}.
\label{Pxy}
\eeq
Analogously, {the fact that $q_y\propto \partial_s T$, together with the relationship $\lambda_0\propto p/\nu$,} allows us to define a (reduced) non-Newtonian thermal conductivity coefficient $\lambda^*(\alpha, a)$  by
\beq
q_y=-\lambda_0\lambda^*(\alpha, a)\frac{\partial T}{\partial y}.
\label{qs}
\eeq
Equations \eqref{Pxy} and \eqref{qs} can be seen as generalizations of Newton's and Fourier's law, equations \eqref{NS2} and \eqref{NS5}, respectively, in the sense that the reduced transport coefficients $\eta^*$ and $\lambda^*$ are {non-linear} functions of the shear rate $a$ and thus they differ from the NS coefficients $\eta^*_\NS$ and $\lambda^*_\NS$ of a granular gas \citep{BDKS98}.
It is important to note that,  due to the coupling between collisional cooling  and gradients in steady states \citep{BC98,SGD04}, the generalized transport coefficients do not reduce to the NS ones  in the absence of shearing ($a=0$). In fact,  at equal wall temperatures and in the absence of shearing, an autonomous thermal gradient appears in the system that is controlled by inelasticity only, so that $\lambda^*(\alpha,0)$ differs from the NS quantity $\lambda^*_\text{NS}(\alpha)$.

It is interesting to remark that, {among the  hypotheses (i)--(iv)} described in \S~\ref{intro}, \emph{only} the  $p=\text{const}$ hypothesis is needed in the framework of Grad's set of equations.

Apart from the generalized coefficients $\eta^*$ and $\lambda^*$, departures from Newton's and Fourier's laws are characterized by normal stress differences and a component of the heat flux orthogonal to the thermal gradient. These effects are measured by the (reduced) directional temperatures
\beq
\theta_x(\alpha,a)=\frac{P_{xx}}{p}, \quad \theta_y(\alpha,a)=\frac{P_{yy}}{p},
\label{thetai}
\eeq
and by a cross conductivity coefficient $\phi^*$ defined as
\beq
q_x=\lambda_0\phi^*(\alpha, a)\frac{\partial T}{\partial y}.
\label{qx}
\eeq
Equation \eqref{thetai} is consistent with the fact that the diagonal elements of the pressure tensor (i.e., the normal stresses) are uniform, while equation \eqref{qx} is consistent with $\partial_s q_x=\text{const}$.
The parameters $\theta_x$ and $\theta_y$ account for the distinction between the diagonal elements ($P_{xx}$ and $P_{yy}$) of the pressure tensor from the hydrostatic pressure $p=[P_{xx}+P_{yy}+(d-2)P_{zz}]/d$. Moreover, $\phi^*$ characterizes the presence of a heat flux component $q_x$ induced by the shearing. These three coefficients are clear consequences of the anisotropy of the system created by the shear flow.
Note that, by symmetry, the coefficients $\eta^*$, $\lambda^*$ and $\theta_i$  are even functions of the shear rate $a$, while $\phi^*$ is an odd function.

Inserting equation \eqref{Pxy}  into the {(exact)} energy balance equation \eqref{balance}, it is straightforward to obtain
\beq
\frac{1}{\nu}\frac{\partial q_y}{\partial y}=p \frac{d(d+2)}{d-1} \lambda^*(\alpha,a)\gamma(\alpha,a),
\label{divq}
\eeq
with
\begin{equation}
\label{gamma}
\gamma(\alpha,a)\equiv\frac{d-1}{d(d+2)}\frac{\eta^*(\alpha,a)a^2-\frac{d}{2}\zeta^*(\alpha)}{\lambda^*(\alpha,a)}.
\end{equation}
Using equation \eqref{qs}, equation \eqref{divq} yields
\begin{equation}
\frac{1}{\nu}\frac{\partial }{\partial y }\left(\frac{1}{\nu}\frac{\partial T}{\partial y}\right)=-2m\gamma(\alpha,a).
\label{nuTdy}
\end{equation}

The technical steps needed to derive the transport coefficients $\eta^*$, $\lambda^*$, $\theta_x$, $\theta_y$ and $\phi^*$, as well as the thermal curvature parameter $\gamma$, in the framework of Grad's method are worked out in Appendix \ref{appB}.

In summary, we have shown that Grad's 13-moment method to solve the Boltzmann equation is consistent with the general assumptions made in \S~\ref{intro}. Moreover, explicit expressions for the generalized non-Newtonian transport coefficients are derived. On the other hand, given the approximate character of Grad's method, a more quantitative agreement with computer simulations is not necessarily expected.

\section{Generalized non-Newtonian hydrodynamics}
\label{class}

    \subsection{Basic hypotheses}
    \label{class1}

Sections \ref{secNS} and \ref{grad} show that the exact balance equations \eqref{Pct} and \eqref{balance} allow for a class of base-state solutions characterized by the following features:
\begin{itemize}
\item
(i) the hydrostatic pressure $p$ is uniform,
\item
(ii) the \emph{reduced} shear rate defined by equation \eqref{udy} is uniform,
\item
(iii)
the shear stress $P_{xy}$ is a {non-linear} function of $a$ but is independent of the thermal gradient $\partial_y T$ and
\item
(iv) the heat flux component $q_y$, properly scaled, is linear in the reduced thermal gradient but depends {non-linearly} on the reduced shear rate $a$.
\end{itemize}

As shown before, in the NS description properties (i)--(iv) are a consequence of the constitutive equations themselves, while in the Grad description one only needs to assume point (i) and then the other three points are derived.

{It is important to remark that hypotheses (iii) and (iv) are fully consistent with the Burnett-order constitutive equations in the Couette--Fourier geometry; taking into account the general structure \citep{CC70} of the Burnett contribution to the shear stress, $P_{xy}^{(2)}$, and to the heat flux, $q_y^{(2)}$, it is straightforward to check that $P_{xy}^{(2)}=q_y^{(2)}=0$ if $\nabla_i u_j=\partial_y u_x\delta_{iy}\delta_{jx}$, $\nabla_i T=\partial_y T\delta_{iy}$ and $\nabla_i p=\partial_y p\delta_{iy}$.  }

The aim of this section is to \emph{assume} the validity of hypotheses (i)--(iv) in the \emph{bulk} domain of the system (i.e., outside the boundary layers) and analyze the different classes of base states that are compatible with them. In doing so, we are assuming that the Boltzmann equation admits for solutions which, in the bulk domain of the system, are essentially in agreement with (i)--(iv), beyond the NS or Grad's approximations. Previous results obtained for ordinary \citep{GS03} and granular \citep{TTMGSD01} gases support the above expectation.

Assumptions (iii) and (iv) can be made more explicit by Eqs.\ \eqref{Pxy} and \eqref{qs}, respectively, where the generalized transport coefficients $\eta^*(\alpha,a)$ and $\lambda^*(\alpha,a)$ have not necessarily the explicit forms provided by Grad's solution. The same can be said about equations \eqref{thetai} and \eqref{qx}. Moreover,  from the energy balance equation \eqref{balance} one can again derive equations \eqref{divq}--\eqref{nuTdy}, provided that the possible spatial dependence of the ratio $\zeta^*\equiv \zeta/\nu$ due to higher-order gradients is discarded. This assumption is supported by kinetic theory calculations \citep{BDKS98} and simulations \citep{TTMGSD01,AS05}.

According to the assumption $p=nT=\text{const}$, the collision frequency defined by equation \eqref{nu0} has the explicit form
\beq
\nu= {K T^{-1/2},\quad K\equiv \frac{p\sigma^{d+1}}{\sqrt{m}c_\eta\Lambda_d}},
\label{nu}
\eeq
and thus equation \eqref{divq} implies that the product $T^{1/2}\partial_y q_y$ is uniform. Moreover, the sign of $\partial_y q_y$ is determined by that of the coefficient $\gamma$. Equivalently, in view of equation \eqref{nuTdy},  the parameter $\gamma$  has a direct  influence on the curvature of the thermal gradient.

{We see from equation \eqref{gamma} that the main} difference between $\gamma$ for elastic and inelastic gases is {the absence or presence of} the  term proportional to $\zeta^*$, {respectively}. {In both cases (i.e., $\zeta^*=0$ or $\zeta^*>0$)},  $\gamma$ is constant. On the other hand, while $\gamma$ is positive definite in the elastic case, its sign results from the competition between viscous heating ($\eta^* a^2$) and inelastic cooling ($d\zeta^*/2$) in the inelastic case. As a consequence, as we will show below, inelasticity spans a more general set of solutions, which includes the elastic profiles as special cases \citep{VU09}.

\subsection{Properties of the hydrodynamic profiles}

{In terms of the scaled spatial variable $s$ defined by equation \eqref{ds},} equations \eqref{udy} and \eqref{nuTdy} take the following forms
\begin{equation}
\frac{\partial u_x}{\partial s}=a,
\label{uds}
\end{equation}
\begin{equation}
\frac{\partial^2 T}{\partial s^2}=-2m\gamma(\alpha,a).
\label{Tds}
\end{equation}
{}From equations \eqref{uds} and \eqref{Tds}, it is straightforward to obtain analytical solutions, in terms of the scaled variable:
\beq
u_x(s)=as+C,
\label{us}
\eeq
\beq
T(s)=-m\gamma(\alpha,a) s^2+As+B,
\label{Ts}
\eeq
where $A$, $B$, $C$ are integration constants. {Please note that integration of the differential equations (\ref{uds}) and (\ref{Tds}) is done independently of the nature of the boundary conditions}. We may set $C=0$ by a Galilean transformation.
The constants $B$ and $A$ represent the values of $T$ and $\partial_s T$, respectively, at a reference point $s=0$. Therefore, {since it is always possible to choose the point $s=0$ within the physical region, henceforth  we can take $B>0$ without loss of generality}.
Note that equations \eqref{us} and \eqref{Ts} imply that $T$ is also \emph{quadratic} when expressed as a function of $u_x$ or, equivalently,
\begin{equation}
\frac{\partial^2 T}{\partial u_x^2}=-2m\frac{\gamma(\alpha,a)}{a^2}.
\label{Tdu}
\end{equation}

Taking into account the definition of $s$ and {equation \eqref{nu}} (with $K=\text{const}$), we may write the derivative $\partial_y^2 T$  in the natural variable $y$ in terms of $\partial_s T$ and $\partial_s^2 T$ as
\beq
\frac{\partial^2 T}{\partial y^2}= K^2 T^{-1/2} \frac{\partial }{\partial s}\left( T^{-1/2}\frac{\partial T}{\partial s}\right)=K^2 T^{-2}\left[ T\frac{\partial^2 T}{\partial s^2}-\frac{1}{2}\left(\frac{\partial T}{\partial s}\right)^2\right].
\label{ecder}
\eeq
By using equation \eqref{Ts}, one gets
\begin{equation}
\frac{\partial^2 T}{\partial y^2}=K^2 T^{-2}\Phi(\alpha,a),
\label{Tdy}
\end{equation}
where {$\Phi$ is also uniform and is defined by}
\begin{equation}
\Phi(\alpha,a)\equiv -2m B\gamma (\alpha,a)-\frac{1}{2}A^2 \label{Phi}.
\end{equation}
In the same spirit as in equation \eqref{Tdu}, the parameter $\Phi$ can be conveniently expressed as
\beq
T\frac{\partial^2 T}{\partial u_x^2}-\frac{1}{2}\left(\frac{\partial T}{\partial u_x}\right)^2=\frac{\Phi(\alpha,a)}{a^2}.
\label{Phibis}
\eeq

In contrast to $\gamma$, the quantity $\Phi$, which measures directly the curvature of the thermal profile, is determined not only by the shear rate and the inelasticity, but also by the temperature boundary conditions through $B$ and $A$.
Similarly, from the identity $\partial_y T=K T^{-1/2}\partial_s T$ and equation \eqref{Ts}, it is straightforward to obtain
\beq
T\left(\frac{\partial T}{\partial y}\right)^2=-2K^2(\Phi+2mT\gamma).
\label{ZZ20}
\eeq
This implies that $\Phi$ is upper bounded: $\Phi\leq -2mT\gamma$. For $\gamma>0$, one has $\Phi\leq -2mT_{\text{max}}\gamma$, while $\Phi\leq 2mT_{\text{min}}|\gamma|$ for $\gamma<0$. Here, $T_{\text{max}}$ and $T_{\text{min}}$ are the maximum and minimum values, respectively, of the  temperature in the system.
Another interesting consequence of equation \eqref{ZZ20} is that, according to the constitutive equation \eqref{qs}, $q_y^2$ is a \emph{linear} function of $T$:
\beq
q_y^2=-\frac{d^2(d+2)^2}{2(d-1)^2}\frac{p^2{\lambda^*}^2}{m^2}(\Phi+2mT\gamma).
\label{qy2}
\eeq
{The same relationship is obtained for $q_x^2$, except that $\lambda^*$ is replaced by $\phi^*$.}

{Since both $\gamma$ and $\Phi$ are constant across the system, equations \eqref{Tdu} and \eqref{Tdy} imply that neither $T(u_x)$ nor $T(y)$ exhibit a  curvature change, i.e., they do not possess an inflection point. On the other hand, this is not necessarily so for the velocity profile $u_x(y)$. To clarify this point, note that, according to equations \eqref{udy} and \eqref{nu},}
\beq
{\frac{\partial^2 u_x}{\partial y^2}=-\frac{Ka}{2}T^{-3/2} \frac{\partial T}{\partial y}.}
\label{ud2y}
\eeq
{Thus (assuming $a>0$), $u_x(y)$ is  convex (concave) in the spatial regions where the temperature increases (decreases). In case the temperature presents a minimum or a maximum at a certain point inside the system, the flow velocity presents there an inflection point. In the derivation of equation \eqref{ud2y} no use of the form of the temperature profile has been made. On the other hand, taking  derivatives on both sides of equation \eqref{ud2y} and using equations \eqref{Tdy} and \eqref{ZZ20}, one obtains }
\beq
{\frac{\partial^3 u_x}{\partial y^3}=-K^3 a T^{-7/2}\left(2\Phi+3m T\gamma\right).}
\label{ud3y}
\eeq
{Therefore, similarly to $T(\partial T/\partial y)^2$ and $q_i^2$, $T^{7/2}\partial^3 u_x/\partial y^3$ is a linear function of temperature.}

Equations \eqref{uds}--{\eqref{ud3y}}  also apply in the NS hydrodynamic description \citep{VU09}, except that $\eta^*(\alpha,a)$, $\lambda^*(\alpha,a)$ and $\gamma(\alpha,a)$ are replaced by their NS counterparts $\eta_\text{NS}^*(\alpha)$, $\lambda_\text{NS}^*(\alpha)$ and $\gamma_\NS(\alpha,a)$, respectively {(see \S~\ref{secNS})}. While $\eta_\text{NS}^*(\alpha)$ and $\lambda_\text{NS}^*(\alpha)$ are independent of the shear rate,   one sees from equation \eqref{gammaNS} that $\gamma_{\text{NS}}(\alpha,a)$ is a linear function of $a^2$.

\subsection{General classification of states}
\label{class.3}

In a previous work \citep{VU09}, the complete set of steady-state solutions based on the signs of the parameters $\gamma$ and $\Phi$ was described in the framework of NS hydrodynamics. It was shown in that work that the analytical expressions of the temperature and flow velocity profiles depend on the signs of these two parameters. Thus, each possible combination of signs of $\gamma$ and $\Phi$ yields a different class of constant pressure laminar flows. Now, we can perform the same analysis in the non-Newtonian regime and  find the \textit{same} set of classes of steady base states.

It is convenient to define the following constants
\beq
  T_0\equiv \frac{|\Phi|}{2m|\gamma|},\quad w^2\equiv \frac{|\Phi|}{2m^2\gamma^2},\quad \ell_0\equiv\frac{wT_0^{1/2}}{2K},\quad s_0\equiv \frac{A}{2m\gamma}.
  \label{ZZ2}
  \eeq
As we will see below, the constants $T_0$, $w$ and $\ell_0$ set the natural scales for $T$, $u_x$ and $y$, respectively.
According to the signs of $\gamma$ and $\Phi$, the following cases are possible:
\begin{itemize}

  \item[\textbf{(1)}] $\gamma>0$.

  This case {[see equation \eqref{gamma}]} corresponds to states where  viscous heating is larger than collisional cooling. Therefore, this class exists only in the presence of shearing ($a\neq 0$) and inelasticity is not required \citep{TTMGSD01}. Note that, according to equation \eqref{Phi}, $\gamma>0$ implies
  \beq
  \Phi<0.
  \label{Phi_XTu}
  \eeq
  {}From equations \eqref{Tds} and \eqref{Tdu}, $T(s)$ and, equivalently, $T(u_x)$ are \emph{convex}. We will refer to this class as XTu. Also, from {equations \eqref{Tdy} and \eqref{Phi_XTu} we conclude that} the profile $T(y)$ is convex as well. Moreover, equation \eqref{qy2} shows that {$q_i^2$ ($i=x,y$)} decreases with increasing temperature.

  Making use of the definitions \eqref{ZZ2} in equation \eqref{Ts}, the quadratic function $T(s)$ can be written as
  \beq
  T(s)=T_0\left[1-\left(\frac{s-s_0}{w}\right)^2\right].
  \label{ZZ1}
  \eeq
 Since $\mathrm{d}y=K^{-1}T^{1/2}\mathrm{d}s$, the relationship between the true and scaled space variables is
      \beq
 y=y_0+ \ell_0\left[\frac{s-s_0}{w}\sqrt{1-\left(\frac{s-s_0}{w}\right)^2}+\sin^{-1}\frac{s-s_0}{w}\right].
  \label{ZZ21}
  \eeq
    {}Eliminating $s$ between equations \eqref{ZZ1} and {\eqref{ZZ21}} one gets $T(y)$ in \emph{implicit} form:
  \beq
  |y-y_0|=\ell_0\left|\sqrt{\frac{T}{T_0}\left(1-\frac{T}{T_0}\right)}+\sin^{-1}\sqrt{1-\frac{T}{T_0}}\right|.
  \label{ZZ8}
  \eeq
Equation \eqref{ZZ21} also provides the velocity profile $u_x(y)$ in implicit form just by replacing $s$ by $u_x/a$:

    \beq
  y=y_0+\ell_0\left[\frac{u_x-u_0}{aw}\sqrt{1-\left(\frac{u_x-u_0}{aw}\right)^2}+\sin^{-1}\frac{u_x-u_0}{aw}\right],
  \label{ZZu}
  \eeq
  where $u_0\equiv as_0$. {A similar replacement in equation \eqref{ZZ1} yields $T$ as a function of $u_x$.}

In the above equations $s_0$ and $y_0$ denote the point where the temperature reaches its maximum value $T=T_0$. This point may be inside the system (i.e., $|y_0|\leq h/2$) or outside the system. In the latter case, the maximum corresponds to a continuation of $T(y)$ into the external region $|y_0|> h/2$. The physical condition $T(y)>0$ implies the domains
\beq
|s-s_0|\leq w,\quad |y-y_0|\leq \frac{\pi}{2}\ell_0.
\label{ZZ18}
\eeq

Although the hydrodynamic profiles in terms of the $s$ variable are quite simple [see equations \eqref{us} and \eqref{Ts}], equations \eqref{ZZ8} and \eqref{ZZu} show that the dependence of $T$ and $u_x$ on the real space variable $y$ is highly nonlinear. A similar comment applies to the cases discussed below (except in the cases LTu and LTy, where the profiles are simpler).

  \item[\textbf{(2)}]
  $\gamma=0$.

  Now viscous heating exactly equals collisional cooling. As a consequence, $T(s)$ and $T(u_x)$ are \emph{linear} functions. For this reason, we formerly referred to this class as LTu \citep{SGV09,VSG10,VGS11}. Moreover, the heat flux is uniform {[see equation \eqref{divq}]}.

  Two possibilities for $\Phi$ are found:
  \begin{itemize}
    \item[\textbf{(2.a)}]
    $\Phi<0$.

    {}From equation \eqref{Phi}, $A^2=2|\Phi|\neq 0$ and the profiles are
      \beq
  T(s)=As+B,
  \label{ZZ3}
  \eeq
 \beq
  u_x(y)=\frac{a}{A}\left[\frac{3}{2}AK(y-{\widetilde{y}_0})\right]^{2/3}-\frac{aB}{A},
  \label{ZZ13}
  \eeq
    \beq
  T(y)=\left[\frac{3}{2}AK(y-{\widetilde{y}_0})\right]^{2/3}.
  \label{ZZ14}
  \eeq
Here ${\widetilde{y}_0}$ represents the mathematical point where $T(y)\to 0$. Obviously, positivity of $T(y)$ requires $y>{\widetilde{y}_0}$ if $A>0$ and $y<{\widetilde{y}_0}$ if $A<0$.
It is possible to prove that equation \eqref{ZZ8} reduces to equation \eqref{ZZ14} in the limit $\gamma\to 0$.

   Notice that, from equation \eqref{gamma},  $\gamma(\alpha,a)=0$ is fulfilled for a threshold shear rate $a_{\text{LTu}}(\alpha)$, whose specific value (for a given $\alpha$) requires the knowledge of $\eta^*$ and $\zeta^*$. In the special case of elastic collisions ($\alpha=1$, i.e., $\zeta^*=0$), $\gamma=0$ implies $a^*_{\text{LTu}}=0$. This corresponds to the conventional Fourier flow of an ordinary gas.
    \item[\textbf{(2.b)}]
    $\Phi=0$.

    This implies $A=0$, so the temperature is uniform and the heat flux vanishes. In this case $s$ is a linear function of $y$ and thus equation \eqref{us} yields
    \beq
    {u_x(y)=a\nu y}
    \label{USF}
    \eeq
{with $\nu=\text{const}$.}
    This state is the well-known uniform (or simple) shear flow  \citep[USF; see, for instance, work by][]{C89}. Note  that here the USF is not generated by the usual Lees--Edwards boundary conditions \citep*{LE72} but by thermal walls in relative motion. The USF needs again the condition $a=a_{\text{LTu}}(\alpha)$. Notice that $\alpha=1$ gives only the trivial equilibrium state of an elastic gas.
  \end{itemize}
  \item[\textbf{(3)}]
  $\gamma<0$.

  In this wide class, inelastic cooling overcomes viscous heating. Therefore, collisions must be inelastic and shearing is not required \citep{BC98}. A negative $\gamma$ implies a \emph{concave} curvature of $T(s)$ and $T(u_x)$, {$q_i^2$ being} an increasing (linear) function of $T$.
  According to equation \eqref{Phi}, we find now three possibilities for the curvature of the temperature profile $T(y)$:
   \begin{itemize}
    \item[\textbf{(3.a)}]
    $\Phi<0$.

    In this subclass, henceforth referred to as {CTu}/XTy, $T(y)$ is a \emph{convex} function.  The profiles are
       \beq
  T(s)=T_0\left[\left(\frac{s-s_0}{w}\right)^2-1\right],
  \label{ZZ4}
  \eeq
     \beq
  y=y_0+\ell_0\left[\frac{s-s_0}{w}\sqrt{\left(\frac{s-s_0}{w}\right)^2-1}
  -\ln\left(\frac{s-s_0}{w}+\sqrt{\left(\frac{s-s_0}{w}\right)^2-1}\right)+\frac{\pi}{2}
  \right],
  \label{ZZ11}
  \eeq
  \beq
  |y-y_0|=\ell_0\left|\sqrt{\frac{T}{T_0}\left(1+\frac{T}{T_0}\right)}-\ln\left(\sqrt{\frac{T}{T_0}}+\sqrt{1+\frac{T}{T_0}}\right)+\frac{\pi}{2}\right|.
  \label{ZZ12}
  \eeq
  In equations \eqref{ZZ4}--\eqref{ZZ12} $s_0$ and $y_0$ denote the \emph{mathematical} point where the temperature reaches its formal minimum value $T=-T_0$. This point must obviously lie outside the system (i.e., $|y_0|> h/2$). The physical condition $T(y)>0$ implies that
\beq
|s-s_0|\geq w,\quad |y-y_0|\geq \frac{\pi}{2}\ell_0.
\label{ZZ19}
\eeq

    \item[\textbf{(3.b)}]
    $\Phi=0$.

   This case corresponds to a \emph{linear} function $T(y)$. Thus, we call this class LTy. From equation \eqref{Phi} we have $B=A^2/4m|\gamma|$ and the profiles are simply
   \beq
  T(s)=m|\gamma|(s-{\widetilde{s}_{0}})^2,
  \label{ZZ6}
  \eeq
   \beq
  u_x(y)=a\left[{\widetilde{s}_{0}}+\left(\frac{2K}{\sqrt{m|\gamma|}}\right)^{1/2}(y-{\widetilde{y}_{0}})^{1/2}\right],
  \label{ZZ15}
  \eeq
  \beq
  T(y)=2K\sqrt{m|\gamma|}(y-\widetilde{y}_{0}),
  \label{ZZ16}
  \eeq
where, without loss of generality, we have assumed $T(h/2)\geq T(-h/2)$.
  Similarly to the LTu case, ${\widetilde{s}_{0}}$ and ${\widetilde{y}_{0}}$ represent the point where $T\to 0$. Thus, one must have $y>{\widetilde{y}_{0}}$.
It is straightforward to reobtain equation \eqref{ZZ16} from equation \eqref{ZZ12} in the limit $\Phi\to 0$.
Note that in the LTy class of states  {$q_i^2/T$} is constant [see equation \eqref{qy2}].

If we denote by
\beq
\delta T^*\equiv{ \frac{1}{K\sqrt{m}}  \frac{\Delta T}{h}}
\label{ZZ27}
\eeq
the \emph{reduced} applied gradient, where $\Delta T\equiv {T({h/2})-T(-{h/2})}$, then the LTy flow requires a transitional value given by
\beq
\delta T^*_{\text{LTy}}(\alpha,a)=2\sqrt{|\gamma(\alpha,a)|}.
\label{ZZ28}
\eeq
Note that, because of expected temperature jumps at the walls \citep{L96,GHW07,N11}, $T(\pm h/2)\neq T_{\pm}$. Moreover, by $T(\pm h/2)$ here we mean the extrapolation to $y=\pm h/2$ of the \emph{bulk} temperature profile, which might differ from the respective temperatures of the fluid layers adjacent to the walls, due to boundary-layer effects.

As we will show below, if $\gamma<0$, {$|\gamma|$ always {increases with decreasing shear rate $a$}, and thus $\delta T^*_{\text{LTy}}(\alpha,a)$ has an upper bound at $a=0$} given by
\beq
\delta T^*_{\text{LTy}}(\alpha,a)\leq 2\sqrt{|\gamma(\alpha,0)|}.
\label{ZZ29}
\eeq
The  LTy state  has been studied previously \citep{BCMR01,BKR09,BKD11,BKD12} in the absence of shearing ($a=0$).

In equation \eqref{ZZ28} it is implicitly assumed that the shear rate $a$ is a free parameter. Reciprocally, given an imposed gradient $\delta T^*\leq2\sqrt{|\gamma(\alpha,0)|}$, it is always possible to find a certain value of the reduced shear rate,
  $a_{\text{LTy}}(\alpha,\delta T^*)$,
 such that
  \beq
 \gamma(\alpha,a_{\text{LTy}}(\alpha,\delta T^*))=-\frac{1}{4}\left(\delta T^*\right)^2.
  \label{ZZ31}
  \eeq

Since $|\gamma|$  is a decreasing function of  $a$, it is obvious that $a_{\text{LTy}}$ increases with decreasing $\delta T^*$. Therefore, the  maximum value occurs at $\delta T^*=0$ (i.e., $\gamma=0$), which coincides with $a_\text{LTu}$ (see figure \ref{LTyLTu}). In other words,
\beq
a_{\text{LTy}}(\alpha,\delta T^*)\leq a_{\text{LTu}}(\alpha).
\label{ZZ32}
\eeq
In fact, the case $a_\text{LTy}=a_\text{LTu}$ corresponds to the USF state.

    \item[\textbf{(3.c)}]
    $\Phi>0$.

    In this  class, $T(y)$ is a \emph{concave} function and so we call this class CTy. The resulting profiles are
  \beq
  T(s)=T_0\left[1+\left(\frac{s-s_0}{w}\right)^2\right],
  \label{ZZ5}
  \eeq
  \beq
 y=y_0+\ell_0\left[\frac{s-s_0}{w}\sqrt{1+\left(\frac{s-s_0}{w}\right)^2}+\sinh^{-1}\frac{s-s_0}{w}\right].
 \label{ZZ9}
 \eeq
  \beq
  |y-y_0|=\ell_0\left|\sqrt{\frac{T}{T_0}\left(\frac{T}{T_0}-1\right)}+\sinh^{-1}\sqrt{\frac{T}{T_0}-1}\right|,
  \label{ZZ10}
  \eeq
 where $s_0$ and $y_0$ denote the point where the temperature reaches its minimum value $T=T_0$.
    \end{itemize}
\end{itemize}

\begin{table}
\begin{center}
\def~{\hphantom{0}}
\begin{tabular}{ccccccccc}
Label&$\text{sign}(\gamma)$&$\text{sign}(\Phi)$&Shearing&Inelasticity&$T(s), T(u_x)$&$T(y)$&{$q_x^2(T)$,$q_y^2(T)$}\\
&&&needed?&needed?&&&\\[3pt]
XTu&$+$&$-$&Yes&No&Convex &Convex &Decreasing\\
LTu&$0$&$-$&Yes$^*$&Yes$^*$&Linear&Convex &Constant\\
USF (LTu)&$0$&$0$&Yes$^\dag$&Yes$^\dag$&Constant &Constant& Zero\\
CTu/XTy&$-$&$-$&No&Yes&Concave &Convex &Increasing\\
LTy&$-$&$0$&No&Yes&Concave& Linear &Increasing\\
CTy&$-$&$+$&No&Yes&Concave &Concave &Increasing\\
\multicolumn{8}{l}{$^*$Except for the Fourier flow of an ordinary gas ($a=0$, $\alpha=1$).}\\
\multicolumn{8}{l}{$^\dag$Except for the equilibrium state of an ordinary gas ($a=0$, $\delta T^*=0$, $\alpha=1$).}\\
\end{tabular}
\caption{Classification of Couette--Fourier flows (see text).}
\label{gF}
\end{center}
\end{table}

\begin{figure}
\begin{center}
\includegraphics[width=.5 \columnwidth]{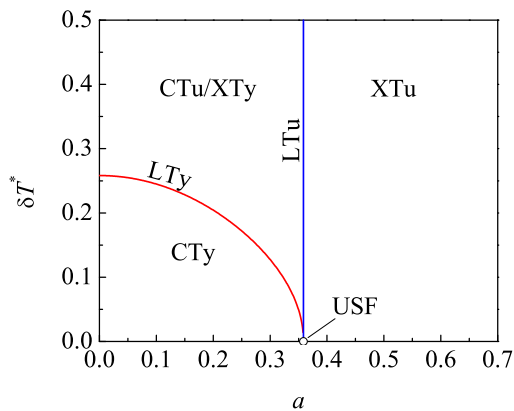}
\end{center}
\caption{Phase diagram illustrating the classification of Couette--Fourier flows. This particular case corresponds to $\alpha=0.9$ and $d=3$,  as obtained from Grad's solution.}
\label{diagram2}
\end{figure}

The main features of the six classes of flows described above are summarized in table \ref{gF}. {Note that these six profile types have been obtained independently of the specific details of the boundary conditions. Once they are specified \citep[and they can be described more realistically than we do later in the simulations, see for instance the work by][]{NAAJS99}, they will determine, for a given value of the coefficient of restitution and in the hydrodynamic bulk (i.e., the region where our four hypotheses {(i)--(iv)}  hold), which type of profile among those in (\ref{ZZ1})--(\ref{ZZ10}) the system will show}.

An illustration of the phase diagram in the $a$-$\delta T^*$ plane at a given value of $\alpha<1$ is presented in figure {\ref{diagram2}}. In fact, the LTu and LTy curves have been obtained from Grad's solution  of the Boltzmann equation (see {\S~\ref{grad}}) for $\alpha=0.9$. It is apparent that the LTy class cannot be attained if $\delta T^*$ is larger than $2\sqrt{|\gamma(\alpha,0)|}$ ({$\simeq 0.26$} in the case displayed in figure \ref{diagram2}) or $a$ is larger than $a_\text{LTu}(\alpha)$ ({$\simeq 0.36$} in the case displayed in figure \ref{diagram2}). As the coefficient of restitution increases, both $|\gamma(\alpha,0)|$ and $a_\text{LTu}$ decrease, so that the CTu/XTy and CTy regions shrink. Of course, in the elastic case only the region XTu persists. All these features are clearly seen in the full phase diagram {depicted in figure} \ref{LTyLTu}.

An interesting remark in the case of \emph{symmetric} walls, i.e., $\delta T^*=0$, is the impossibility of having a temperature profile that is concave in the variables $s$ or $u_x$ but convex in the variable $y$ (CTu/XTy region). As figure \ref{diagram2} shows, if $\delta T^*=0$ and both plates are at rest ($a=0$), $T(y)$ is concave. As shearing is introduced and increased, the concavities of {$T(y)$} and $T(u_x)$ decrease until  the value $a=a_\text{LTu}$ is reached, where the temperature is uniform and $u_x(y)$ is linear (USF). Further increase of the shearing produces  convex profiles $T(y)$ and $T(u_x)$. Thus, the existence of the `hybrid' CTu/XTy region requires  asymmetric walls ($\delta T^*\neq 0$).

\section{Comparison with computer simulations}
\label{results}

\subsection{Simulation details}

In this section we present the results obtained from DSMC and MD simulations for hard spheres ($d=3$) and compare them with the analytical results derived from Grad's theory. The simulation methods that we used for DSMC and MD simulations are similar to those in our previous works and have been explained in detail elsewhere {\citep*{LVU09,VU09,VGS11,VSG11}}. We will briefly recall that  DSMC yields an exact numerical solution of the corresponding kinetic equation (inelastic Boltzmann equation in this case), whereas MD yields a solution of the equations of motion of the particles. Therefore, the main difference between results from both methods is that MD simulations lack the bias of the inherent statistical approximation of the Boltzmann equation, where velocity correlations between particles which are about to collide are not considered. As in our previous work \citep{VGS11},  the global solid volume fraction in the MD simulations has been taken equal to $7\times 10^{-3}$ (dilute limit), using $N\sim 10^4$--$10^5$ particles. In DSMC simulations we take a similar number of particles, $N=2\times 10^5$. The boundary conditions used here are analogous {in} both methods. When a particle collides with a wall, its velocity is updated following the rule $\boldsymbol {v}\rightarrow\boldsymbol{v}'+U_\pm{{\boldsymbol{e}_x}}$. The first {contribution} ($\boldsymbol{v}'$) of the new particle velocity is due to thermal boundary condition, {while} the second {contribution} ($U_\pm{{\boldsymbol{e}_x}}$) is due to wall {motion}. The horizontal components of $\boldsymbol{v}'$ are randomly {drawn} from a Maxwellian distribution (at a temperature $T_\pm$), whereas the normal component {$v_y'$ is sampled} from a Rayleigh probability distribution: {$P(|v_y'|)=(m|v_y'|/T_{\pm})e^{-m{v_y'}^2/2T_\pm}$} \citep{AG97}.

At a given value of $\alpha$, we consider a common wall distance $h=15(\sqrt{2}\pi \overline{n}\sigma^2)^{-1}$, where $\overline{n}$ is the average density, and  8 different series of simulations with $T_+/T_-=2.5$, $5.0$, $7.5$, \ldots, $20.0$. For each value of the wall temperature ratio, a number of wall velocity differences $(U_+-U_-)/\sqrt{2T_-/m}\approx 2$--$20$ is taken.

Once the steady state is reached, the local values of  $p(y)$, $u_x(y)$, $T(y)$ and $\nu(y)\propto p(y)[T(y)]^{-1/2}$ are coarse-grained into $25$ layers \citep{VSG11}. The local shear rate $a$ is obtained from  equation \eqref{udy}. Next, the local curvature parameters $\gamma$ and $\Phi$ are obtained from equations \eqref{Tdu} and \eqref{Phibis}, respectively. In order to evaluate the derivatives $\partial u_x/\partial y$, $\partial T/\partial u_x$ and $\partial^2 T/\partial u_x^2$, the profiles $u_x(y)$ and $T(u_x)$ are fitted to polynomials (typically of fifth degree).

\begin{table}
\begin{center}
\def~{\hphantom{0}}
\begin{tabular}{cccccccc}
System&$\frac{U_+-U_-}{\sqrt{2T_-/m}}$&$\frac{T(-h/2)}{T_-}$&$\frac{T(h/2)}{T_-}$
&$\frac{u_x(-h/2)-U_-}{\sqrt{2T_-/m}}$&$\frac{U_+-u_x(h/2)}{\sqrt{2T_-/m}}$&$\frac{n(-h/2)}{\overline{n}}$&$\frac{n(h/2)}{\overline{n}}$\\[3pt]
A&$5.5$&$0.9706$&$7.1799$&$	0.1892$&$	0.6080$&$	2.1357$&$	0.2939$\\
{B}&${10.6}$&${1.2953}$&${8.9035}$&${0.2634}$&${0.8651}$&${2.8482}$&${0.4193}$\\
{C}&${11.3}$&${1.3397}$&${9.2022}$&${0.2715}$&${0.8457}$&${3.0905}$&${0.4610}$\\
D&$11.85$&$1.3741$&$9.3722$&$0.2727$&$0.8584$&$	3.1788$&$	0.4897$\\
E&$14.0$&$1.5154$&$10.2501$&$0.2861$&$0.8934$&$	3.5821$&$	0.5625$\\
F&$17.0$&$1.7316$&$10.9953$&$0.3062$&$0.9302$&$	4.1538$&$	0.6889$\\
\end{tabular}
\caption{Values of the wall velocity difference and of the hydrodynamic fields near the walls for six representative systems. In all the cases  $\alpha=0.9$, $h=15(\sqrt{2}\pi \overline{n}\sigma^2)^{-1}$ and $T_+/T_-=10$.}
\label{cases1}
\end{center}
\end{table}

\begin{table}
\begin{center}
\def~{\hphantom{0}}
\begin{tabular}{ccccccc}
System&$K$&$\delta T^*$&$a$&$\gamma$&$\Phi$&Class\\[3pt]
A&$ {0.994}$&$0.1589$&$0.2491$&$-0.0110$&$~0.0218$&CTy\\
{B}&${1.044}$&${0.1064}$&${0.3597}$&${-0.0022}$&${-0.00004}$&LTy\\
{C}&${1.038}$&${0.1008}$&${0.3697}$&${-0.0013}$&${-0.0049}$&CTu/XTy\\
D&${1.047}$&$0.0972$&$0.3753$&$-0.0006$&$-0.0087$&LTu\\
E&${1.062}$&$0.0854$&$0.3994$&$~0.0017$&$-0.0260$&XTu\\
F&${1.073}$&$0.0683$&$0.4251$&$~0.0044$&$-0.0558$&XTu\\
\end{tabular}
\caption{Values of the parameters $K$ [equation \eqref{nu}], $\delta T^*$ [equation \eqref{ZZ27}], $a$ [equation \eqref{udy}], $\gamma$ [equation \eqref{nuTdy}] and $\Phi$ [equation \eqref{Tdy}] for the systems described in table \protect\ref{cases1}. The right-most column shows the class each system belongs to.}
\label{cases2}
\end{center}
\end{table}

\begin{figure}
\begin{center}
\includegraphics[width=\columnwidth]{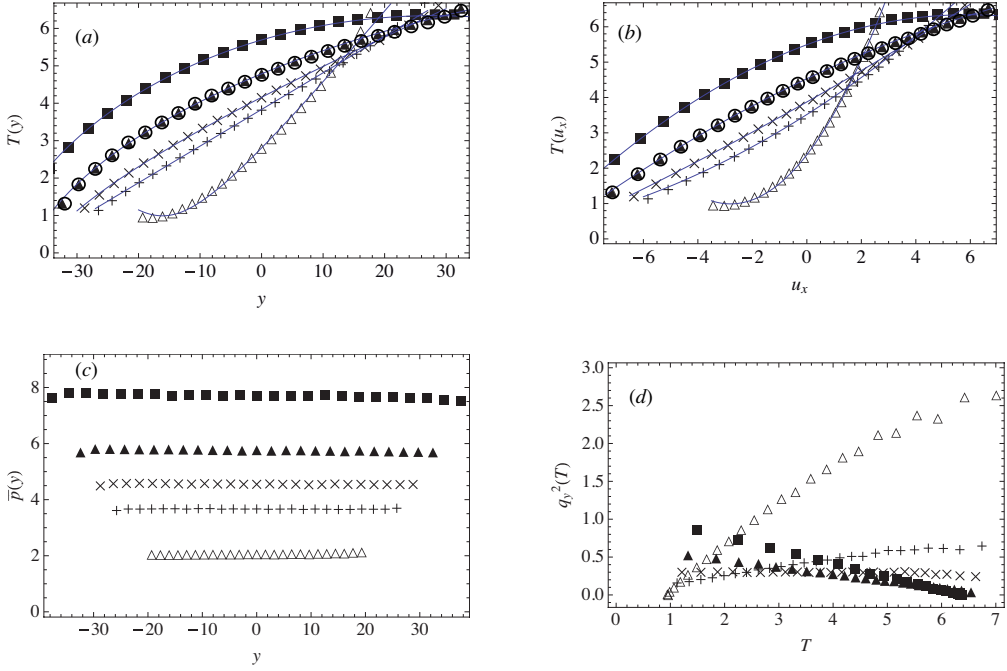}
\end{center}
\caption{ (\textit{a}) Profile $T(y)$, (\textit{b}) parametric plot $T(u_x)$, (\textit{c}) profile $\overline{p}(y)\equiv (n_r/\overline{n})p(y)$ and (\textit{d}) parametric plot $q_y^2(T)$, as obtained from DSMC simulations for the systems A ($\triangle$), B ($+$), D ($\times$), E ($\blacktriangle$) and F ($\blacksquare$) described in table \protect\ref{cases1}. Lines in (\emph{a}) and (\emph{b}) represent the theoretical profiles. {Additionally, we present $T(y)$ and $T(u_x)$ plots {($\bigcirc$)} as obtained from MD simulations for state E.} {The quantities are scaled with respect to the reference units described in the text.}} \label{profiles1}
\end{figure}

\begin{figure}
\begin{center}
\includegraphics[width=\columnwidth]{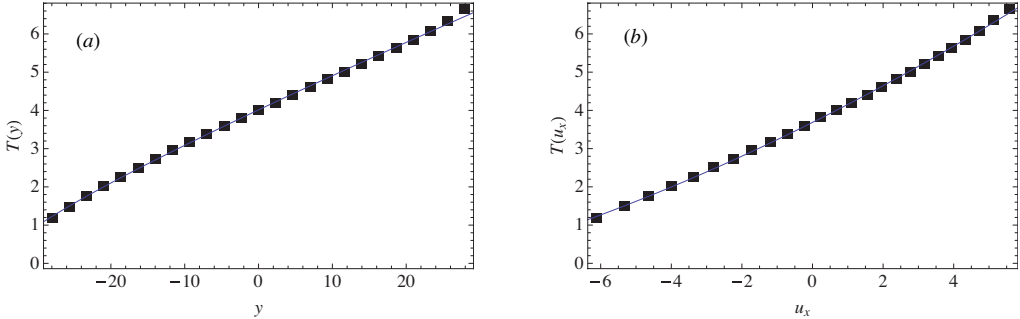}
\end{center}
\caption{(\textit{a}) Profile $T(y)$ and (\textit{b}) parametric plot $T(u_x)$, as obtained from DSMC simulations for the system C described in table \protect\ref{cases1}. Lines  represent the theoretical profiles. {The quantities are scaled with respect to the reference units described in the text.}} \label{profiles2}
\end{figure}


\subsection{Hydrodynamic profiles}

{Similarly to previous works, we have observed in all simulation runs that $p$, $a$, $\gamma$ and $\Phi$ practically remain constant in the central layers of the system}. {Thus, in the subsequent analysis the local values of $p$, $a$, $\gamma$ and $\Phi$ are replaced by \emph{global} values obtained by a spatial average in the bulk domain.}

{The five} classes of flows summarized in table \ref{gF} and figure \ref{diagram2} are found in the simulations. {The USF state with thermal walls, which requires $\delta T^*=0$,  was analyzed elsewhere \citep{VSG10,VGS11} and is not considered here}. As an illustration,  let us consider the six representative systems  described in table \ref{cases1}. We observe that, at fixed values $h=15(\sqrt{2}\pi \overline{n}\sigma^2)^{-1}$ and $T_+/T_-=10$, the fluid temperatures near the walls do not coincide with the imposed wall {values (temperature jumps)}.
As we increase shearing, the differences $T(\pm h/2)-T_\pm$ increase, changing from negative to positive values (see three first columns in table \ref{cases1}). As for the velocity slips \citep{L96}, i.e., the differences $u_x(\pm h/2)- U_\pm$, they {also tend to increase (with one exception) with increasing shearing.}

In what follows, as in former works \citep{VU09,VGS11}, we take the quantities near the cold wall  as reference units. Thus, $m$, $T_r\equiv T(-h/2)$ and $\tau_r\equiv 1/\nu(-h/2)$ define the units of mass, energy and time, respectively. Therefore, distances are measured in units of the nominal mean free path
{$\tau_r\sqrt{T_r/m}=5c_\eta/(16\sqrt{\pi}n_r\sigma^2)$}, where $n_r\equiv n(-h/2)$. Moreover, the density is scaled with respect to $n_r$.
The steady-state hydrodynamic profiles for the systems in table \ref{cases1} are shown in figures \ref{profiles1} and \ref{profiles2}. Since the  profiles in system C are very close to those of systems B and D, {system C} is absent in figure \ref{profiles1} and its temperature profiles are shown separately in figure \ref{profiles2}.
It is quite apparent that the pressure is {practically} uniform in all the cases, thus confirming the hypothesis (i) made in \S~\ref{class}. Notice also that, even though in the simulations  the size is fixed at $h=15(\sqrt{2}\pi \overline{n}\sigma^2)^{-1}$, the dimensionless size of systems A--E in the units of our choice varies since $n_r/\overline{n}$ is different in each case. Moreover, in our reduced units $p(y)\approx 1$ at all places and systems and so, for better visualization, in figure \ref{profiles1}(\textit{c}) we choose to plot $\overline{p}(y)$ instead. The (bulk) temperature
profile $T(y)$ is concave for system A, linear for system B and convex for systems C--F. Regarding the profile $T(u_x)$, it is concave for systems A--C, linear for system D and convex for systems E and F. The parametric dependence of $q_y^2$ versus $T$ is linear (in the bulk region) in all the cases, in agreement with equation \eqref{qy2}, being {an increasing function} for systems A--C, constant for system D and decreasing for systems E and F.

The values of the quantities $K$, $\delta T^*$, $a$, $\gamma$ and $\Phi$ obtained from the hydrodynamic profiles of systems A--F are displayed in table \ref{cases2}.
Notice in this table that the measured values of $\Phi$ and $\gamma$ correctly predict in all cases the observed curvatures of $T(y)$ and $T(u_x)$, respectively. Moreover, we have obtained a very close approach to LTy and LTu states in {systems} B and D (for which {$\Phi=-0.00004$} and $\gamma=-0.0006$. respectively).

We introduced the {simulation} values of $K$,  $a$, $\gamma$ and $\Phi$ into the, according to our description,  corresponding theoretical profiles for $T(y)$ and $T(u_x)$, {by using the pertinent (depending on the signs of $\gamma$ and $\Phi$)} expressions given in \S~\ref{class.3}.
{It is worth remarking that the theoretical profiles $T(y)$ do  not depend on the separate values of  $K$,  $a$, $\gamma$ and $\Phi$ but only on the two combinations $T_0$ and $\ell_0$ [cf.\ equations \eqref{ZZ2}]; as for the theoretical profiles  $T(u_x)$, they depend on the same parameter $T_0$ as before plus the combination $aw$.}
The resulting profiles are included in figures \ref{profiles1}(\emph{a}), \ref{profiles1}(\emph{b}) and \ref{profiles2}, where the integration constants $y_0$ and $u_0$ are determined as to reproduce $T$ and $u_x$ at $y=0$.
As we can observe, the agreement between the theoretical curves {from our generalized hydrodynamic description (see \S~\ref{class.3}) and} simulation data is excellent, the deviations typically being restricted to 1-2 layers near the cold wall and 2-4 layers near the hot wall. Those small deviations can be due to boundary-layer effects and/or to residual limitations of the  hydrodynamic description exposed in \S~\ref{class}. In any case, it is worth remarking that the local mean free path (inversely proportional to the local density) is  larger near the hot wall (where deviations present a longer range) than near the cold wall. {As a matter of fact, in the employed reference units, the mean free path is $\sim 1$ near the cold wall and $\sim n(-h/2)/n(h/2)=6$--$7$ near the hot wall.} It is also interesting to note that the lack of agreement near the boundaries seems to become less important as the shearing increases (i.e., from system A to system F).

The simulation data plotted in figures \ref{profiles1} and \ref{profiles2} have been obtained from the DSMC method but they perfectly agree with those obtained from MD. As an example,  we compare the results obtained from both simulation methods in one of the curves of figures \ref{profiles1}(\emph{a}) and \ref{profiles1}(\emph{b}).

\subsection{Transport coefficients}

Once we have checked that the steady base states discussed in \S~\ref{class} are supported by the simulations, we now proceed to present the simulation results for the transport coefficients and compare  them with Grad's theoretical predictions.

\begin{figure}
\begin{center}
\includegraphics[width=\columnwidth]{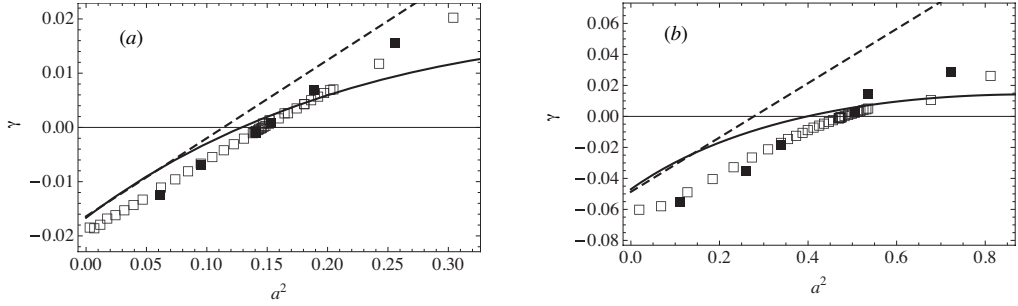}
\end{center}
\caption{Thermal curvature parameter $\gamma$ as a function of shear rate squared $a^2$ for two values of the coefficient of restitution: (\textit{a}) $\alpha=0.9$  and (\textit{b}) $\alpha=0.7$. Lines represent results from Grad's analytical solution (solid lines) and from the NS prediction (dashed lines), while symbols stand for DSMC ($\square$) and MD ($\blacksquare$) simulations.} \label{G1}
\end{figure}

As a general trend, we have observed a relatively good {semi-quantitative} agreement between simulation and {Grad's} theory for all relevant quantities, except for the reduced thermal conductivity $\lambda^*$ and for the reduced viscosity $\eta^*$ at low $a$. In figure \ref{G1} we plot the results for the thermal curvature parameter $\gamma$ for two different values of the coefficient of restitution: $\alpha=0.9$ and $0.7$. We detect, both in simulations and theory, the aforementioned transition from $\gamma<0$ for low shear rates to $\gamma>0$ for higher shear rates. This transition is also predicted by the NS solution \citep{VU09}, in which case $\gamma$ is a linear function of $a^2$ {[see equation \eqref{gammaNS}]}. As we see, the true parameter $\gamma$ has a more complex dependency on $a$. It is apparent that Grad's theory predicts well the value $a=a_\text{LTu}$ where $\gamma=0$, as already shown elsewhere \citep{VSG10,VGS11}. It is also noteworthy that, in the region $\gamma>0$, Grad's theory does a better job for $\alpha=0.7$ than for $\alpha=0.9$.
It might seem surprising that both NS and Grad's predictions for $\gamma$ show significant discrepancies with simulation data in the region of small shear rates, especially for $\alpha=0.7$. The explanation lies in the fact that, apart from  $a$ and $\delta T^*$,  $\gamma$ is an additional measure of the strength of the gradients, which in the limit $a\to 0$ is governed by $\alpha$ and thus cannot be done arbitrarily small for finite inelasticity.

\begin{figure}
\begin{center}
\includegraphics[width=\columnwidth]{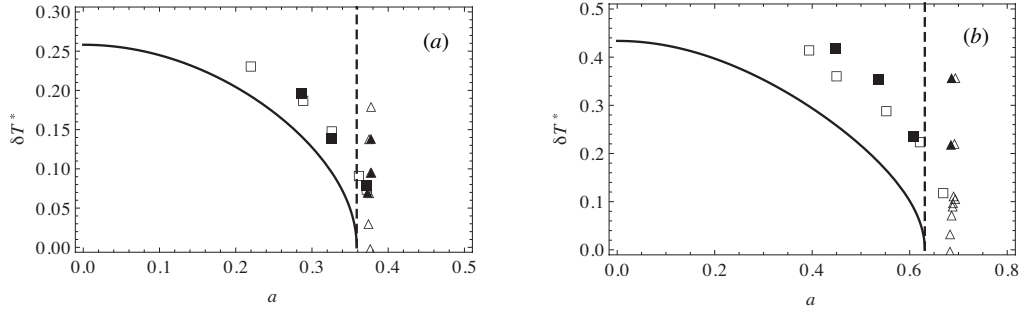}
\end{center}
\caption{Phase diagram in the plane $\delta T^*$ vs $a$ (see figures \ref{LTyLTu} and \ref{diagram2}) for two values of $\alpha$: (\textit{a}) $\alpha=0.9$ and (\textit{b}) $\alpha=0.7$. Lines stand for the analytical solution from Grad's method, while open and solid symbols stand for DSMC and MD simulations, respectively. The LTy curve is represented by  solid lines (theory) and  squares (simulation), while the LTu line is represented by dashed lines (theory) and  triangles (simulation). } \label{LTy1}
\end{figure}

\begin{figure}
\begin{center}
\includegraphics[width=.5\columnwidth]{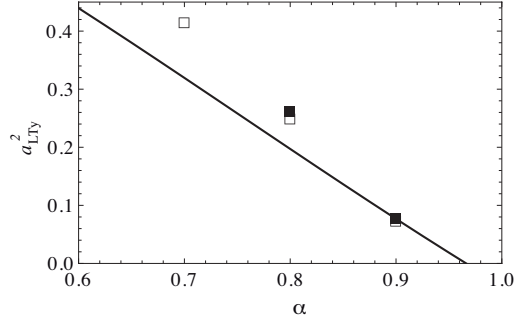}
\end{center}
\caption{Threshold value $a_{\text{LTy}}^2$ for which the linear $T(y)$ occurs if $\delta T^*=0.015$, as a function of the coefficient of restitution.  Line stands for Grad's method solution, while open and solid symbols stand for DSMC and MD simulations, respectively.}  \label{LTy2}
\end{figure}

As discussed in \S~\ref{class.3}, for  a given value of $\alpha$, it is possible to find pairs $(\delta T^*,a)$ such that the temperature profile $T(y)$ is linear (LTy states).  It is also possible to find a value of $a$, independent of $\delta T^*$, where the temperature profile $T(u_x)$ is linear (LTu states). These two loci split the plane $\delta T^*$ vs $a$ into the three regions sketched in figure \ref{diagram2}. We represent in figure \ref{LTy1} the phase diagram, as obtained from our simulations, for  (\textit{a}) $\alpha=0.9$ and (\textit{b}) $\alpha=0.7$. For comparison, the curves predicted by Grad's solution are also included. As we see, the agreement between theory and simulation is qualitatively good for both values of $\alpha$. As a complement, figure \ref{LTy2} shows the threshold value $a_{\text{LTy}}^2$ versus the coefficient of restitution for  $\delta T^*=0.015$. We observe that  the LTy is not possible for this value of the slope $\delta T^*$ if $\alpha\ge 0.967$.

\begin{figure}
\begin{center}
\includegraphics[width=\columnwidth]{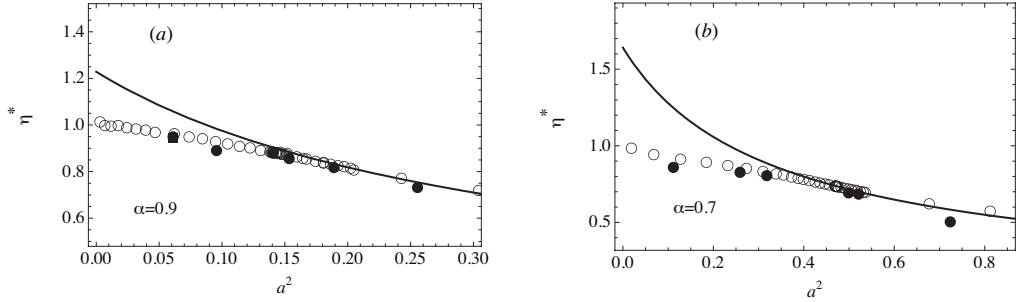}
\end{center}
\caption{Generalized viscosity $\eta^*$ as a function of $a^2$ for (\textit{a}) $\alpha=0.9$ and (\textit{b}) $\alpha=0.7$.  Lines stand for Grad's method solution, while open and solid symbols stand for DSMC and MD simulations, respectively.} \label{G2a}
\end{figure}

\begin{figure}
\begin{center}
\includegraphics[width=\columnwidth]{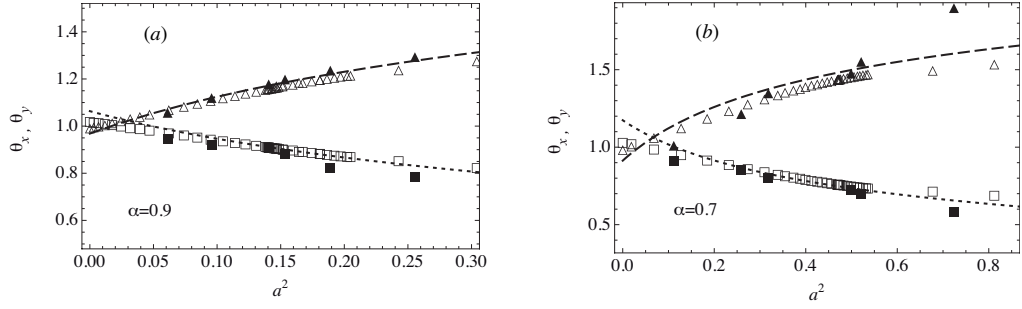}
\end{center}
\caption{Reduced normal stress components $\theta_x$ (dashed lines and triangles) and $\theta_y$ (dotted lines and squares)  as  functions of $a^2$ for (\textit{a}) $\alpha=0.9$ and (\textit{b}) $\alpha=0.7$.  Lines stand for Grad's method solution, while open and solid symbols stand for DSMC and MD simulations, respectively.} \label{G2b}
\end{figure}

In figures \ref{G2a} and \ref{G2b} we plot the shear-rate dependence of the reduced shear viscosity $\eta^*$ and of the normalized diagonal components of the stress tensor $\theta_x$ and $\theta_y$, respectively. It is quite apparent that, except for the shear viscosity in the range of low shear rates, the agreement between Grad's analytical solution and DSMC and MD simulations is {quite good} ({somewhat better  for $\alpha=0.9$}). {The agreement is specially good {around} the LTu states (i.e., $a^2\approx 0.15$ and $a^2\approx 0.55$ for $\alpha=0.9$ and $\alpha=0.7$, respectively), as previously reported \citep{VSG10,VGS11}. Figure \ref{G2a} shows that the non-linear shear viscosity decreases with increasing shear rate (`shear thinning' effect). In what concerns the reduced directional temperatures, figure \ref{G2b} shows that $\theta_x$ ($\theta_y$) increases (decreases) with increasing shearing. It is interesting to note that $\theta_x<\theta_y$ for very small shear rates, until both quantities cross at a certain value of $a$. This phenomenon is qualitatively captured by Grad's solution.}
Comparison between figures \ref{G2a}(\emph{a}) and  \ref{G2a}(\emph{b})  shows that, as the inelasticity decreases, the region of shear rates corresponding to $\gamma<0$, and hence the region with worse Grad's predictions, shrinks. In fact, in the purely elastic case ($\alpha=1$) the Grad expression for $\eta^*$ is rather accurate \citep{GS03}.

\begin{figure}
\begin{center}
\includegraphics[width=\columnwidth]{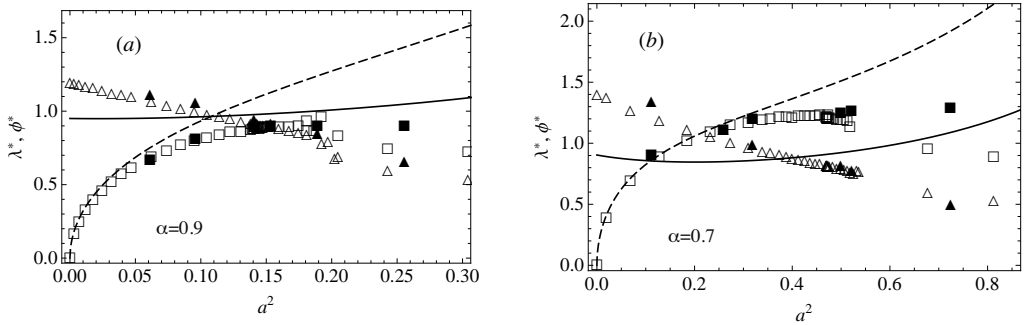}
\end{center}
\caption{Generalized  thermal conductivity $\lambda^*$ (solid lines and triangles) and heat flux cross coefficient $\phi^*$ (dashed lines and squares) as  functions of $a^2$ for (\textit{a}) $\alpha=0.9$ and (\textit{b}) $\alpha=0.7$.  Lines stand for Grad's method solution, while open and solid symbols stand for DSMC and MD simulations, respectively.} \label{G3}
\end{figure}

Finally, in figure \ref{G3} we plot the results for the two heat flux transport coefficients ({thermal conductivity $\lambda^*$ and cross coefficient $\phi^*$}). As {already explained}, there is in general a {(non-Newtonian)} horizontal component of the heat flux,  from which the cross thermal conductivity coefficient $\phi^*$ results. Perhaps surprisingly, we {find} that the agreement between Grad's theory and simulations is better for the cross coefficient $\phi^*$ than for the thermal conductivity $\lambda^*$. {Moreover,  while} Grad's theory predicts that $\lambda^*$  {weakly increases with $a$ ($\alpha=0.9$) or exhibits a non-monotonic behavior ($\alpha=0.7$), simulations yield {a decreasing} $\lambda^*$ vs $a$}. On the contrary, the agreement for the cross coefficient is {qualitatively good}, since $\phi^*$ vs $a$ is increasing both for Grad's theory and simulation. This agreement is very good in the region of low shear rates up to the threshold value for LTu states (as expected), whereas for higher shear rates the theory and simulation results tend to separate.

A final comment regarding the comparison between simulation and Grad's theory is in order. According to equation \eqref{gamma}, $\zeta^*\propto k\eta^* a^2-\lambda^*\gamma$, where $k\equiv (d-1)/d(d+2)$. Since the reduced cooling rate $\zeta^*$ is satisfactorily captured by Grad's method [(see equation \eqref{3.7}], we conclude that the deviations of $\eta^*$, $\lambda^*$ and $\gamma$ from the simulation data are not entirely independent and are somewhat constrained by the {combination $\frac{2}{15}\eta^* a^2-\lambda^*\gamma$ (note that $k=\frac{2}{15}$ for $d=3$)}. In fact, figures \ref{G1}, \ref{G2a} and \ref{G3} show that, in the region with $\gamma<0$, $|\gamma|$ and $\lambda^*$ are underestimated by Grad's solution, while $\eta^*$ is overestimated. In the region of $\gamma>0$, however, $\eta^*$ is quite accurate, so that the underestimation of $\gamma$ is compensated by an overestimation of $\lambda^*$.
It is interesting to remark that the accuracy of Grad's quantitative predictions is highly correlated with the magnitude of the thermal curvature parameter $\gamma$, i.e., the smaller $|\gamma|$ the better the general performance of Grad's solution. In fact, the agreement between theory and simulation is quite good in the LTu state ($\gamma=0$), as previously shown by \cite{VSG10,VGS11}. This confirms the role played by $\gamma$ as an intrinsic measure of the strength of the gradients \citep{VU09}.

\section{Conclusions}
\label{conclusions}

\begin{table}
\begin{center}
\def~{\hphantom{0}}
\begin{tabular}{ccccc}
Features&\multicolumn{3}{c}{Level of description}\\
&NS&Grad&Gener. non-Newton. &Simulation\\
[3pt]
(i) $p=\text{const}$&Derived&Assumed&Assumed&Observed\\
(ii) $a=\text{const}$&Derived&Derived&Assumed&Observed\\
(iii) $P_{xy}\neq F(\partial_y T)$&Construction&Derived&Assumed&Observed\\
(iv) $q_{y}\propto \partial_y T$&Construction&Derived&Assumed&Observed\\
$P_{xx}\neq P_{yy}\neq P_{zz}$&No&Yes&Yes&Yes\\
$q_{x}\neq 0$&No&Yes&Yes&Yes\\
Transport coefficients&Explicit&Explicit&Unspecified&Measured\\
\end{tabular}
\caption{
Hypotheses (i)--(iv) and main features of the plane Couette--Fourier flow according to the level of description:
NS (\S\,\protect\ref{secNS}), Grad's 13-moment method (\S\,\ref{grad}), generalized non-Newtonian hydrodynamics (\S\,\ref{class}) and simulation (\S\,\protect\ref{results}).}
\label{tab0}
\end{center}
\end{table}

\subsection{Summary}

We have studied in this paper the laminar flows in a low density granular gas confined between two infinite parallel walls, which, in general, are at different temperatures. Additionally, the granular gas can be sheared if there is relative motion between both walls. We have described a general classification of steady granular Couette--Fourier flows that occur in this system, at constant pressure, for arbitrarily large velocity and temperature gradients. We have shown that, due to symmetries in the system, the steady-state equations for the flow velocity and temperature are quite simple, even in the non-Newtonian regime, and have a straightforward analytical solution. Moreover, the type of solutions for the hydrodynamic profiles turn out to be dependent on just two constant parameters: the thermal curvature {coefficients} $\gamma$ and $\Phi$. The former is proportional to the second derivative of $T$ in a spatial variable scaled with collision frequency, while $\Phi$ is related to the second derivative in the natural spatial variable. Depending on the different possible combinations of signs of these two parameters, the corresponding steady profiles can be grouped into five different classes of flows, each one having peculiar properties (see table \ref{gF}).

The main conclusions of this work are that {the assumptions made on the form of the hydrodynamic profiles [see equations \eqref{udy} and \eqref{Pxy}--\eqref{qx}], as well as} the associated classification of flows, have been validated by three independent routes. From a theoretical perspective, we have obtained an exact solution of the set of moment equations derived from Grad's method applied to the inelastic Boltzmann equation. Next, we have simulated the Couette--Fourier flows by  using the DSMC method (which numerically solves the Boltzmann equation)  and  MD simulations (which numerically solve the equations of motion of the system of inelastic hard spheres).

This triple validation extends in a non-trivial way some of the qualitative features of the NS description to the realm of non-Newtonian hydrodynamics. This is summarized in table \ref{tab0}. As shown in \S~\ref{secNS}, the NS constitutive equations, complemented by the momentum and energy balance equations in the steady Couette--Fourier geometry, imply the fulfillment of points (i)--(iv) without further assumptions. On the other hand, they do not account for normal stress differences or a heat flux component parallel to the flow. This is remedied by Grad's moment method, in which case only  hypothesis (i) on the constancy of pressure is needed. A more general non-Newtonian treatment makes use of the four assumptions on the same footing, thus allowing us to accommodate for any specific form of the generalized transport coefficients. Finally, the simulation results are seen to support {the validity} of those assumptions, providing as well the dependence of the main quantities on both the shear rate and the coefficient of normal restitution.
However, it must be {kept in mind} that, while simulations are essentially consistent to a large extent with the generalized hydrodynamic description of \S~\ref{class}, slight deviations due to the high intricacy of the Boltzmann equation cannot be discarded. Those small deviations have been reported in the case of the pure Fourier flow for elastic hard spheres by \cite*{MASG94}.

{While Grad's moment method supports the four assumptions (i)--(iv), as well as the existence of normal stress differences and a heat flux component orthogonal to the thermal gradient (see table \ref{tab0}), we have observed that a quantitative agreement with simulations  is generally good near the LTu state (i.e., for small values of $|\gamma|$) only.  As the magnitude of the thermal curvature parameter $\gamma$ increases, some transport coefficients ($\eta^*$ for $\gamma>0$, $\phi^*$ for $\gamma<0$ and $\theta_x$ and $\theta_y$ for both $\gamma<0$ and $\gamma>0$) are better predicted by Grad's theory than other ones ($\eta^*$ for $\gamma<0$, $\phi^*$ for $\gamma>0$ and $\lambda^*$ for both $\gamma<0$ and $\gamma>0$).}

\subsection{Discussion}

The  signs of $\gamma$ and $\Phi$ depend on both the physical properties of the granular gas and  the boundary conditions. However, rather than analyzing the interaction between gas and wall, our work is focused, similarly to previous works \citep{VU09,VSG10}, on the {\emph{bulk}} properties of the gas itself and we study all possible transitions between the different classes of flows. {All class transitions} have been generated by using  the usual hard wall boundary conditions, both in DSMC and MD simulations \citep*[see for instance the work by][where the same boundary conditions are used for simulation of thermal walls]{GHW07}. The phase diagram obtained from simulations is completely analogous to the theoretical one, depicted in figures \ref{LTyLTu} and \ref{diagram2}, {as shown in figure \ref{LTy1}}. We have checked in the simulations that, as in figure \ref{LTyLTu}, only two of the {five} possible flow classes (see table \ref{gF}) define surfaces in the three-parameter space $\{\alpha, \delta T^*, a\}$. They divide this space into {three} regions that define {three} other entire classes of granular flows. Thus, we have taken these surfaces as a reference for our analysis of flow class transitions. One of the surfaces is the LTu flow class ($\gamma=0$), characterized by {linear} temperature {vs} flow velocity profiles, and already studied in former works \citep{VSG10,VGS11}. The other surface is the LTy class ($\Phi=0$), characterized by linear temperature vs vertical {coordinate  profiles}. The LTy surface is always below (lower shear rates) the LTu surface (figure \ref{LTyLTu}), except for walls at the same temperature ($\delta T^*=0$ plane), where they coincide, defining a curve that is the remaining sixth flow class, which can be regarded as a subclass of the LTy or LTu classes. This class (or subclass) is actually the well known uniform shear flow (USF), i.e., constant $T$ and linear $u_x(y)$. Note that here the USF is achieved with thermal walls rather than with generalized periodic boundary conditions \citep{LE72}.
Regarding the other classes, the first region (CTy) is below the LTy surface and is characterized by $\gamma<0$ and $\Phi>0$. The second region (CTu/XTy) occupies the space between the LTy and LTu surfaces, being characterized with $\gamma<0$ and $\Phi<0$. Finally, the third region (XTu) is above the LTu surface and corresponds to $\gamma>0$ and $\Phi<0$ (see figure \ref{LTyLTu}).

One important difference between LTy and LTu classes is that, while LTu flows are possible for arbitrarily large $\delta T^*$, the LTy flows are restricted to values of $\delta T^*$ smaller than a threshold value $\delta T^*_\text{LTy}(\alpha,a)$, which has an upper bound at $a=0$  (see figures \ref{diagram2} and \ref{LTy1}). The agreement between theory and simulation in this aspect is qualitatively good. In particular, we have checked that  a too large $\delta T^*$ in the simulations results in a direct LTu transition without passing through {an} LTy transition, when increasing shear rate from $a=0$. For instance, for $\alpha=0.9$, and following results in figure \ref{LTy1}(\emph{a}), a value of {$\delta T^*=0.3$}  suffices to suppress the LTy transition. Thus, in this case we would already start from $\Phi<0$ at $a=0$, never entering the class of flows with concave $T(y)$.

{We have not detected so far instabilities (departures from laminar flows) in the simulations.This is reasonable since the flows that we have analyzed are either below or not  far above the LTu surface, and thus they occur at low Reynolds number \Rey~\citep[LTu flows typically have $\Rey\le 100$, see the work by][]{VGS11}. In order to see higher \Rey~we would need to separate much further above the LTu surface, at extremely large shear rates, or apply larger $\delta T^*$.}

{In conclusion}, we have described in detail, by means of theoretical and computational studies, all possible classes of base laminar flows for a low density granular gas in a Couette--Fourier flow geometry. {Those classes differ in the curvature of the $T(y)$ and $T(u_x)$ profiles but otherwise they can be described within a common framework characterized by a heat flux proportional to the thermal gradient and uniform stress tensor and reduced shear rate. This unified setting encompasses known and new  states, from the Fourier flow of ordinary gases to the uniform shear flow of granular gases, from the symmetric Couette flow of ordinary gases to Fourier-like flows of granular gases with constant thermal gradient and from states with a magnitude of the heat flux $|\boldsymbol{q}|$ increasing with temperature to states with a decreasing, a constant or even a zero $|\boldsymbol{q}|$.}

\subsection{Outlook}

The flow classes described in this work might be useful for future works in a variety of problems in granular dynamics, such as the study of a granular impurity under Couette flow  {\citep{GV10,VSG11}}.
This implies that the same set of flow classes should exist for the granular impurity; LTu and LTy classes for instance.
This may have implications to segregation conditions for  a granular impurity \citep{JY02,GV09,GV10}.  Moreover, a complete determination of the steady base states
is convenient for studies of {instabilities \citep*{HL91,WJS96,AN98,NAAJS99,KM03,ASL08}}.
Furthermore, analogous temperature curvature properties are observed for the same geometry in moderately dense granular gases, except that {for higher densities} a region with temperature curvature inflections grows from the boundaries {\citep{L96,AN98}}.
Thus, we expect some of the conclusions of the present analysis to be useful for instability in quite generic problems of granular flow. We are currently working on extensions of this work in granular segregation and flow instability.

\begin{acknowledgments}
This research  has been supported by the Spanish Government through Grants No.\ {FIS2010-16587} and (only F.V.R.)  No.\ {MAT2009-14351-C02-02}. Partial support from the Junta de Extremadura (Spain) through Grant No.\ GR10158, partially financed by FEDER (Fondo Europeo de Desarrollo Regional) funds, is also acknowledged.
\end{acknowledgments}

\appendix

\section{Navier-–Stokes transport coefficients}
\label{appNS}

The expressions for the NS transport coefficients are \citep{BDKS98,BC01}
\beq
\eta_\NS^*(\alpha)=\frac{1}{\beta_1(\alpha)+\frac{1}{2}\zeta^*(\alpha)},
\label{etaNS}
\eeq
\beq
\kappa_\NS^*(\alpha)=\frac{1}{\beta_2(\alpha)-\frac{2d}{d-1}\zeta^*(\alpha)},
\label{kappaNS}
\eeq
\beq
\mu_\NS^*(\alpha)=\frac{\frac{d}{d-1}\zeta^*(\alpha)}{\left[\beta_2(\alpha)-\frac{2d}{d-1}\zeta^*(\alpha)\right]
\left[\beta_2(\alpha)-\frac{3d}{2(d-1)}\zeta^*(\alpha)\right]}.
\label{muNS}
\eeq
Here,
\begin{equation}
\label{3.6}
\beta_1(\alpha)=\frac{1+\alpha}{2}\left[1-\frac{d-1}{2d}(1-\alpha)\right],
\end{equation}
\begin{equation}
\label{3.8}
\beta_2(\alpha)=\frac{1+\alpha}{2}\left[1+\frac{3}{8}\frac{d+8}{d-1}(1-\alpha)\right],
\end{equation}
and the reduced cooling rate $\zeta^*(\alpha)$ is given by equation \eqref{3.7}.
In equations \eqref{3.7} and \eqref{etaNS}--\eqref{3.8}, terms associated with the deviation of the homogeneous cooling state distribution from a Maxwellian have been neglected {\citep*{GSM07}}.

\section{Explicit expressions in Grad's approximation}
\label{appB}

Taking into account in equations \eqref{3.9}--\eqref{3.13}  the form of the fluxes given by equations  \eqref{Pxy}--\eqref{qx}, one gets, after some algebra,
\begin{equation}
\label{4}
a\left[\theta_y-(\beta_1+\zeta^*)\eta^*\right]-\frac{2d}{d-1}\gamma\phi^* =0,
\end{equation}
\begin{equation}
\label{5}
(\beta_1+\zeta^*)\theta_x-2\eta^*a^2+\frac{2d}{d-1}\gamma\lambda^* =\beta_1,
\end{equation}
\begin{equation}
\label{6}
(\beta_1+\zeta^*)\theta_y+\frac{6d}{d-1}\gamma\lambda^* =\beta_1,
\end{equation}
\begin{equation}
\label{7}
(d+4)a\left[\eta^*+\frac{d}{d-1}\lambda^*\right]-(d+2)\beta_2\phi^*=0,
\end{equation}
\begin{equation}
\label{8}
\frac{d+4}{2}\theta_y-\frac{d+2}{2}\beta_2\lambda^*+\frac{d}{d-1}a\phi^*=1.
\end{equation}
The algebraic equations \eqref{4}--\eqref{8} allow one to express $\eta^*$, $\lambda^*$, $\theta_x$, $\theta_y$ and $\phi^*$ in terms of $a$, $\alpha$ and $\gamma$ as

\begin{eqnarray}
\label{13}
\eta^*&=&\Delta^{-1}\left\{2d^2(d+4){\beta}_{1}a^2
-(d-1)^2(d+2)^2{\beta}_{1}\beta_{2}^2+2d\left[d(d+4)\left((d+4)\beta_{1}-2\overline{\beta}_{1}\right)\right.\right.
\nn
&&\left.\left.-6(d-1)(d+2)
\beta_{2}\right]\gamma\right\}
,
\end{eqnarray}
\begin{equation}
\label{14}
\lambda^*=\Delta^{-1}(d-1)\left\{\left[
2\overline{\beta}_{1}-(d+4)\beta_{1}\right]\left[(d-1)(d+2)\overline{\beta}_{1}\beta_{2}
+2d(d+4)\gamma\right]-2d(d+4)\beta_{1}a^2\right\},
\end{equation}
\begin{eqnarray}
\label{11}
\theta_{x}&=&(\Delta\overline{\beta}_{1})^{-1}
\Big\{\beta_{1}(2a^2+\overline{\beta}_{1}^2)\left[2d^2(d+4)a^2-
(d-1)^2(d+2)^2\beta_{2}^2\right]+2 d
\left[2d(d+4)a^2\right.\nonumber\\
& &\left.\times \left((d+2)\beta_{1}-2\overline{\beta}_{1}\right)
-(d-1)(d+2)\beta_{2}
\left(12a^2+
\overline{\beta}_{1}\left(2\overline{\beta}_{1}+3(d+4)\beta_{1}\right)\right)
\right]\gamma
\nn
&&
-8d^2(d+4)\left[\overline{\beta}_{1}+(d+4)\beta_{1}\right]\gamma^2\Big\},
\end{eqnarray}
\begin{eqnarray}
\label{12}
\theta_{y}&=&\Delta^{-1}\Big\{2d^2(d+4)\overline{\beta}_{1}\beta_{1}a^2
-\left[(d-1)(d+2){\beta}_{1}\beta_{2}+12d\gamma\right]
\nonumber\\
& &  \times
\left[(d-1)(d+2)\overline{\beta}_{1}\beta_{2}+2d(d+4)\gamma\right]
\Big\},
\end{eqnarray}
\begin{equation}
\label{15}
\phi^*=\Delta^{-1}(d-1)(d+4)a\left\{d\overline{\beta}_{1}\left[2
\overline{\beta}_{1}-(d+4){\beta}_{1}\right]-
(d-1)(d+2){\beta}_{1}\beta_{2}-12d\gamma
\right\},
\end{equation}
where $\overline{\beta}_{1}\equiv\beta_{1}+\zeta^*$ and
\begin{eqnarray}
\label{16}
\Delta&\equiv&2d^2(d+4)(\overline{\beta}_{1}^2
-6\gamma)a^2-(d-1)^2(d+2)^2
\overline{\beta}_{1}^2\beta_{2}^2
-8d(d-1)(d+2)(d+4)\overline{\beta}_{1}
\beta_{2}\gamma
\nonumber\\
& &
-12d^2(d+4)^2\gamma^2.
\end{eqnarray}
Finally, substitution of $\eta^*$ and $\lambda^*$ into equation \eqref{gamma} yields a quadratic equation for $\gamma$. Its physical solution gives $\gamma$ as a function of the shear rate $a$ and the coefficient of restitution $\alpha$.

Setting $\gamma=0$ in equations \eqref{gamma}, \eqref{13} and \eqref{14}, we get the prediction for the LTu threshold shear rate in Grad's approximation. The result is
\beq
a_{\text{LTu}}(\alpha)=\sqrt{\frac{d\zeta^*}{2\beta_1}}\overline{\beta}_1.
\eeq
The expressions for the LTu transport coefficients $\eta^*$, $\lambda^*$, $\theta_x$, $\theta_y$ and $\phi^*$ are obtained by making $a=a_\text{Ltu}$ and $\gamma=0$ in equations \eqref{13}--\eqref{16}. The explicit expressions have been given elsewhere \citep{VGS11}.

In the absence of shearing ($a\to 0$), equations \eqref{gamma} and \eqref{13}--\eqref{16} yield
\beq
\gamma^*=-\frac{(d+2)(d-1)\beta_2\zeta^*\overline{\beta}_1}{2\left[(d+2)^2\beta_1+(3d^2+10d-4)\zeta^*\right]},
\eeq
\beqa
\eta^*&=&\frac{(d+2)^2\beta_1+(3d^2+10d-4)\zeta^*}{(d+2)^2\beta_1+2(d^2+3d-2)\zeta^*}\Big\{
\beta_1\left[(d+2)^2(d-1)\beta_2+d^2(d+4)\zeta^*\right]
\nn
&&
+d\zeta^*\left[3(d+2)(d-1)\beta_2+d(d+4)\zeta^*\right]\Big\}/
{(d-1)(d+2)^2\beta_2\overline{\beta}_1^2},
\eeqa
\beq
\lambda^*=\frac{(d+2)^2\beta_1+(3d^2+10d-4)\zeta^*}{(d+2)^2\beta_2\overline{\beta}_1},
\eeq
\beq
\theta_x=\frac{(d+2)\beta_1+d\zeta^*}{(d+2)\overline{\beta}_1},
\eeq
\beq
\theta_y=\frac{(d+2)\beta_1+3d\zeta^*}{(d+2)\overline{\beta}_1},
\eeq
\beqa
\frac{\phi^*}{a}&=&\frac{d+4}{(d-1)(d+2)^3\beta_2^2\overline{\beta}_1^2}\frac{(d+2)^2\beta_1+(3d^2+10d-4)\zeta^*}{(d+2)^2\beta_1+2(d^2+3d-2)\zeta^*}
\Big\{d(d+2)^2\beta_1^2+\beta_1\left[(d+2)^2\right.
\nn
&&\left.\times (d-1)\beta_2 +2d^2(2d+7)\zeta^*\right]+d\zeta^*\left[3(d+2)(d-1)\beta_2+(3d^2-10d+4)\zeta^*\right]\Big\}.\nn
&&
\eeqa
In the elastic case ($\zeta^*\to 0$, $\beta_1\to 1$, $\beta_2\to 1$), one has $\theta_x\to 1$, $\theta_y\to 1$, $\lambda^*\to 1$, $\eta^*\to 1$, $\gamma\to 0$ and $\phi^*/a\to (2d-1)(d+4)/(d-1)(d+2)$, which corresponds to the Fourier flow of conventional gases.






\end{document}